\documentclass[12pt]{article}

\pdfoutput=0
\usepackage{amsmath}
\usepackage{amsfonts}
\usepackage{amssymb}
\usepackage{float}
\usepackage{ulem}
\usepackage{graphicx}

\usepackage{color}
\restylefloat{table}

\newlength{\abstractwidth}

\usepackage{epsf}
\usepackage{cancel}

\renewcommand{\thanks}[1]{\footnote{#1}}

\renewcommand{\theequation}{\thesection.\arabic{equation}}
\newcommand{\bea}{\begin{eqnarray}}
\newcommand{\eea}{\end{eqnarray}}
\newcommand{\ee}{\end{equation}}
\newcommand{\be}{\begin{equation}}

\newcommand{\ea}{\end{array}}
\newcommand{\bac}{\begin{array}{c}}
\newcommand{\bacc}{\begin{array}{cc}}
\newcommand{\barcl}{\begin{array}{r@{}c@{}l}}
\newcommand{\brcl}{\begin{array}{rcl}}
\newcommand{\bdm}{\begin{displaymath}}
\newcommand{\edm}{\end{displaymath}}

\def\cN{{\cal N}}

\def\no{\nonumber}

\def\eqn#1{eq.~(\ref{#1})}

\def\nvec5{{\tilde n}}

\def\id{\protect{{1 \kern-.28em {\rm l}}}}

\def\spa#1.#2{\left\langle#1\,#2\right\rangle}
\def\spb#1.#2{\left[#1\,#2\right]}

\newcommand{\ha}{{\hat a}}

\usepackage{color}

\renewcommand{\theequation}{2.\arabic{equation}}
\begin{document}

\textwidth 170mm
\textheight 230mm
\topmargin -1cm
\oddsidemargin-1cm \evensidemargin -1cm
\topskip 9mm
\headsep9pt

\overfullrule=0pt
\parskip=2pt
\parindent=12pt
\headheight=0in \headsep=0in \topmargin=0in \oddsidemargin=0in

\vspace{ -3cm} \thispagestyle{empty} \vspace{-1cm}

\begin{flushright}
UUITP-62/18 \ \ \
NORDITA 2018-130
\end{flushright}

\vspace{ -0.1cm}

\begin{center}

{\Large \bf Non-Abelian gauged supergravities as double copies}

\bigskip

{\large Marco Chiodaroli,${}^{a}$ Murat G\"{u}naydin,${}^{b,c}$ Henrik Johansson,${}^{a,d}$ \\ and Radu Roiban${}^{b}$}

\medskip

\smallskip
${}^a${ Department of Physics and Astronomy, Uppsala University,
	SE-75108 Uppsala, Sweden\\
}
\smallskip
${}^b${ Institute for Gravitation and the Cosmos \\
 The Pennsylvania State University, University Park PA 16802, USA \\
}
\smallskip
${}^c$ {Stanford Institute for Theoretical Physics and Department of Physics, Stanford University, Stanford,
CA 94305, USA } \\
 \smallskip
${}^d${ Nordita, Stockholm University and KTH Royal Institute of Technology, \\
Roslagstullsbacken 23, SE-10691 Stockholm, Sweden\\
}
\bigskip

\end{center}

\begin{abstract}

Scattering amplitudes have the potential to provide new insights to the study of supergravity theories with gauged R-symmetry and Minkowski vacua. Such gaugings break supersymmetry spontaneously, either partly or completely.
In this paper, we develop a framework for double-copy constructions of  Abelian and non-Abelian gaugings of
${\cal N} = 8$ supergravity with these properties.
They are  generally obtained as the double copy of a spontaneously-broken (possibly supersymmeric) gauge theory and a theory
with explicitly-broken supersymmetry.
We first identify purely-adjoint deformations of ${\cal N} = 4$ super-Yang-Mills theory that preserve the duality between color and kinematics. A combination of Higgsing and orbifolding yields the needed duality-satisfying gauge-theory factors with multiple matter representations.
We present three explicit examples. Two are Cremmer-Scherk-Schwarz  gaugings with unbroken ${\cal N} = 6, 4$ supersymmetry and $U(1)$ gauge group. The third has unbroken ${\cal N} = 4$ supersymmetry and $SU(2) \times U(1)$ gauge group. We also discuss examples in which the double-copy method gives theories with explicitly-broken supersymmetry.

\end{abstract}

\baselineskip=16pt
\setcounter{equation}{0}
\setcounter{footnote}{0}

\newpage

\tableofcontents

\newpage

\section{Introduction and summary of results}
\renewcommand{\theequation}{1.\arabic{equation}}
\setcounter{equation}{0}

Gauged supergravities -- supergravities in which all or part of the R-symmetry is gauged and some fields transform
nontrivially under the gauge group -- have been the subject of active investigation  both in their own right and because of their
relation to the low-energy limit of string compactifications with fluxes.
Such theories typically feature a rich array of interesting physical properties. Due to the presence of a nontrivial potential
for scalar fields, they can allow for a non-vanishing cosmological constant, moduli stabilization, and spontaneous breaking of supersymmetry.
Moreover, gauged supergravities with Anti-de Sitter vacua play a prominent role in the low-energy limit of the
holographic relation between string and gauge theories.

An $SO(8)$ gauging of four-dimensional $\cN=8$ supergravity 
 \cite{Cremmer:1979up}
 was first formulated by  de Wit and Nicolai in ref. \cite{deWit:1982bul}. Non-compact gaugings of $\cN=8$ supergravity via deformations of de Wit and Nicolai construction were later studied in refs.~\cite{Hull:1984vg,Hull:1988jw,Hull:1984wa}. Compact and non-compact gaugings of maximal supergravity in five dimensions were  obtained by one of the current authors and collaborators  \cite{Gunaydin:1984qu}, including a  gauging  which has zero cosmological constant and preserves $\cN=2$ supersymmetry \cite{Gunaydin:1985tb}.
Along similar lines, gaugings of  $\cN=2$ supersymmetric Maxwell-Einstein supergravity theories in five dimensions were first given by Sierra, Townsend and one of the current authors in refs. \cite{Gunaydin:1983bi, Gunaydin:1984ak, Gunaydin:1986fg} and further generalized to theories involving tensor fields in refs. \cite{Gunaydin:2000ph,Gunaydin:2000xk}.

The introduction of the so-called embedding tensor formalism \cite{Nicolai:2000sc,Nicolai:2001sv,deWit:2002vt,deWit:2004nw,
deWit:2005hv}, which
makes use of a manifestly U-duality-covariant formulation of the action, provided novel strategies for the
construction of gauged supergravities.
At the level of the supergravity Lagrangian, gauge covariant  derivatives are written in a U-duality-covariant form
\begin{equation}
\partial_\mu \rightarrow D_\mu \equiv \partial_\mu - g A^M_\mu \Theta_M^{\ \alpha} \  t_\alpha \ ,
\end{equation}
where the index $M$ labels all the vector fields in the theory and $t_\alpha$ are generators of the U-duality group.
The embedding tensor $\Theta_M^{\ \alpha}$  specifies the explicit
embedding of the gauge group into the global symmetry group (U-duality).
The closure of the gauge algebra implies that the tensor $\Theta_M^{\ \alpha}$ must satisfy a quadratic constraint
while supersymmetry implies that it must also satisfy a linear constraint.
 All quantities relevant to the supergravity Lagrangian, including the scalar potential, can then be expressed in terms of $\Theta_M^{\ \alpha}$ (see \cite{Samtleben:2008pe} for a detailed review). The embedding tensor formalism led to the discovery of new families of gaugings of $\cN=8$ supergravity, including a new $SO(8)$ family
 in four dimensions \cite{Catino:2013ppa,DallAgata:2011aa,DallAgata:2012mfj}. Despite this progress, a complete classification of gauged supergravities has thus far remained elusive and is the subject of ongoing efforts.

In recent years, the study of scattering amplitudes has provided a new
perspective on various gravity and supergravity theories.
Particular progress has been achieved by  the double-copy construction introduced by Bern, Carrasco and one
of the current
authors \cite{Bern:2008qj,Bern:2010ue}, which allows the construction of gravitational amplitudes using gauge-theory
building blocks.
 The key ingredient of this construction is an organization of gauge-theory amplitudes in which numerator factors obey the same algebraic conditions as color factors. If presentations of amplitudes with this property are available, the gauge theory is said to obey color/kinematics duality. Amplitudes which are invariant under linearized diffeomorphisms, and thus can be regarded as the amplitudes of a gravitational theory, are then obtained by substituting color factors with a second set of numerators.

It has been shown that many families of gravitational and non-gravitational theories are amenable to double-copy methods.
These include pure supergravities  \cite{Bern:2008qj,Johansson:2014zca,Bern:2011rj}, various finite or infinite families of matter-coupled supergravities \cite{Carrasco:2012ca,Chiodaroli:2013upa,Bern:2013yya,Chiodaroli:2015wal,Johansson:2017bfl,Ben-Shahar:2018uie}, and conformal supergravities \cite{Johansson:2017srf,Johansson:2018ues}.
Effective non-gravitational theories for which  a double-copy construction is known include the  Dirac-Born-Infeld and the special Galileon theories, as well as the nonlinear sigma model \cite{Chen:2013fya, Chen:2014dfa, Cachazo:2014xea,  Du:2016tbc,  Carrasco:2016ldy, Chen:2016zwe, Cheung:2016prv}.
Double-copy structures have also been identified for various string-theory amplitudes, \cite{Stieberger:2009hq,BjerrumBohr:2009rd,BjerrumBohr:2010zs,Mafra:2011kj,Carrasco:2016ygv,He:2015wgf,Mafra:2017ioj,Azevedo:2018dgo,Fu:2018hpu}, within the Cachazo-He-Yuan formalism  \cite{Cachazo:2013hca,Cachazo:2013gna,Cachazo:2013iea,DuTengBCJ,Bjerrum-Bohr:2016axv}, in the context of ambitwistor string theories
\cite{Mason:2013sva}, and at the level of linearized supermultiplets  \cite{Anastasiou:2014qba,Anastasiou:2015vba,Anastasiou:2016csv,Anastasiou:2017nsz,Anastasiou:2018rdx}.

While various Yang-Mills-Einstein theories have been investigated in  detail \cite{Chiodaroli:2014xia,Cachazo:2014xea,Cachazo:2014nsa,Casali:2015vta,Chiodaroli:2016jqw,Nandan:2016pya,Chiodaroli:2017ngp,DuTengBCJ,Teng:2017tbo}, amplitudes in gauged supergravities have been comparatively less studied.
Very recently, the current authors formulated a double-copy construction for simple $U(1)_R$  gauged $\cN=2$ supergravities that admit Minkowski vacua \cite{Chiodaroli:2017ngp}. On general grounds, one can show that such theories have massive gravitini and spontaneously-broken supersymmetry. To capture this property, it is natural to consider a double-copy construction that~\cite{Chiodaroli:2017ngp}:
\begin{enumerate}
\item Contains massive gravitini, which are realized as bilinears of massive $W$ bosons from one gauge theory and  massive fermions from the other;
\item Reproduces the construction for the ungauged supergravity theory (with unbroken supersymmetry) in the massless limit.
\end{enumerate}
Based on these requirements, the desired  double copy must have the schematic form
\begin{eqnarray}
\Big(\text{gauged supergravity} \Big) = \Big( \text{Higgs-YM} \Big) \otimes \Big(\cancel{\rm S}\text{YM}  \Big)\, .
\end{eqnarray}
The two  theories entering the construction are a spontaneously-broken gauge theory and a gauge theory with explicitly-broken supersymmetry.
Both theories are obtained by starting from a higher-dimensional massless theory, which is then taken on the Coulomb branch
as outlined in ref.~\cite{Chiodaroli:2015rdg}.  The second gauge-theory factor
is obtained with an additional orbifolding procedure which results in a theory with massless adjoint bosons and
massive fermions in a matter representation.\footnote{While there are many ways to break supersymmetry explicitly, the one
used in ref.~\cite{Chiodaroli:2015rdg} and outlined here is singled out by the requirement of color/kinematics duality and
by the details of the double-copy with the chosen spontaneously-broken theory.}
Finally, the free parameters in the family of $U(1)_R$ gaugings are identified with the freedom of choosing the vacuum
expectation values (VEVs) in the gauge theories entering the double copy.

In this paper we address two important problems:
(1) the description of non-Abelian R-symmetry gaugings as double copies and
(2) the application of the double-copy method for studying gaugings of $\cN=8$ supergravity.
For non-Abelian gaugings, we will seek a double-copy construction which obeys the two requirement listed above as well as the additional one that:
\begin{enumerate}
	\item[3.] One of the gauge theories contains trilinear scalar couplings, which depend on an anti-symmetric tensor $F^{IJK}$, which in turn determines the non-Abelian part of the R-symmetry gauging.
\end{enumerate}
 Similarly to the simpler construction described in ref.~\cite{Chiodaroli:2017ngp}, the gauge theories entering the double-copy
 construction are obtained, through a combination of Higgsing and orbifolding, from higher-dimensional theories which
 obey color/kinematics duality.
In contrast to our previous work, which has as starting point higher-dimensional pure Yang-Mills (YM) or super-Yang-Mills (SYM) theories,
we begin with massive deformations of these theories that are chosen to preserve the duality between color and kinematics. We show that
color/kinematics duality amounts to requiring that the fermionic mass matrix $M$ squares to a diagonal matrix and obeys
\be
\big[    \{ \Gamma^I,  M  \} , \Gamma^J\, \big] + i \lambda F^{IJK} \Gamma^K = 0   \ ,
\label{masterintro}
\ee
where $\Gamma^I$ are the Dirac matrices in the extra dimensions, unrelated to the four-dimensional Dirac matrices. The
tensor $F^{IJK}$ obeys either  standard or  modified Jacobi identities (and is related to the structure constants of the
supergravity unbroken  gauge group).
Gauge theories solving the constraint \eqref{masterintro} are then taken on the Coulomb branch and subjected to an orbifold projection, in close analogy with the strategy of ref. \cite{Chiodaroli:2017ngp}.\footnote{In case the theory is orbifolded in the unbroken gauge phase, an ungauged supergravity would be obtained.}

We shall focus on three explicit examples. The first two are Cremmer-Scherk-Schwarz (CSS)-type gaugings \cite{Scherk:1978ta,Scherk:1979zr,Cremmer:1979uq} with unbroken Abelian gauge groups and  $\cN=6,4$ residual supersymmetry. The last example is a non-Abelian gauging with $SU(2) \times U(1)$ unbroken gauge group and $\cN=4$ residual supersymmetry. In all these cases, the double copy allows to quickly calculate the mass spectra of the theory and to access information about the unbroken symmetries. Our results are in agreement with
the supergravity literature \cite{DallAgata:2011aa,DallAgata:2012tne,Catino:2013ppa}. We
also present  examples in which the double copy involves two explicitly-broken gauge theories and hence produces a supergravity theory with explicit supersymmetry breaking.

The structure of this paper is as follows. In Section \ref{Sec-CK}, we study
higher-dimensional massive theories which preserve color/kinematics duality, obtaining simple constraints of the form (\ref{masterintro}). In Section \ref{Sec-sol}, we focus on $\cN=4$ SYM theory and list some explicit solutions to the constraints. In Section \ref{Sec-DC}, we spell out the Higgsing and orbifolding procedure employed to construct the lower-dimensional theories which are then used to obtain various gauged supergravities with the double-copy technique, provided that  some simple consistency requirements are satisfied. In Section \ref{Sec-examples}, we present examples of our construction, leaving a complete classification of gaugings with Minkowski vacua to future work. We conclude the paper with a discussion of our results.

\section{Color/kinematics duality with massive fermions and $\phi^3$ interactions \label{Sec-CK}}
\renewcommand{\theequation}{2.\arabic{equation}}
\setcounter{equation}{0}

Our starting point is a massless gauge theory with scalars and fermions in arbitrary dimension $D$, which is general enough
to include the Lagrangians of the gauge theories discussed in ref.~\cite{Chiodaroli:2015wal} as well as certain
supersymmetric theories:
\bea
{\cal L} \!\!\!\! &=& \!\!\!\! - {1 \over 4} (F^{\hat a}_{\mu \nu})^2 + {1 \over 2} (D_\mu \phi^{\hat a I})^2
- {g^2 \over 4} f^{\hat a \hat b \hat e} f^{\hat c \hat d \hat e} \phi^{\hat a I} \phi^{\hat b J} \phi^{\hat c I} \phi^{\hat d J} + {i \over 2} \bar \psi \cancel{D} \psi  + {g \over 2} \phi^{\hat a I} \bar \psi \Gamma^{I} t^{\hat a} \psi \ . \qquad \label{L-undeformed}
\eea
If fermions are taken in definite (possibly reducible) matter representations,\footnote{By matter representation we
mean a representation of the gauge group that is different from the adjoint representation.} this Lagrangian is that of the non-supersymmetric theory entering the
double-copy construction for homogeneous supergravities~\cite{Chiodaroli:2015wal}.
Following a standard construction, $L$-loop $n$-point gauge-theory  amplitudes are written as
\begin{equation}
{\cal A}^{(L)}_{n} = i^{L-1} g^{n-2+2L} \sum_{i \in \text{cubic}}\, \int \frac{d^{LD}\ell}{(2\pi)^{LD}} \frac{1}{S_i} \frac{c_{i} n_i}{D_i} \, ,
\label{gaugepresentation}
\end{equation}
where the sum runs over cubic graphs, $D_i$ denotes the product of
the  inverse scalar propagators of the cubic graph $i$, and $S_i$ are symmetry factors. $c_i$
and $n_i$ are group-theory and kinematic factors associated  with that graph, respectively.
The defining commutation relations of the gauge group as well as its Jacobi identities imply that there exist  triplets of graphs
$\{i,j,k\}$ such that $c_i + c_j + c_k = 0$.
A scattering amplitude is said to obey color/kinematics duality if the kinematic numerators obey the same algebraic relations as the color factors:
\begin{equation}
n_i - n_j = n_k  \quad \Leftrightarrow \quad  c_i - c_j=c_k \, .
\label{duality}
\end{equation}
Imposing color/kinematics duality on the two-fermion-two-scalar amplitudes following from the Lagrangian (\ref{L-undeformed})
constrains  the $\Gamma$ matrices to be  generators of a Clifford algebra \cite{Chiodaroli:2015wal},
\begin{equation}
 \big\{ \Gamma^I, \Gamma^J \big\} = -2 \delta^{IJ} \ ;
\end{equation}
in turn this implies that (\ref{L-undeformed}) can be regarded as a higher-dimensional YM theory with fermions reduced to four-dimensions.
The remaining two parameters -- the dimension $D$ and the choice of irreducible representations for fermions --
 reproduce the existing classification
of homogeneous supergravities~\cite{deWit:1991nm}.
In addition, by allowing the fermions in eq.~(\ref{L-undeformed}) to be in the adjoint representation, the Lagrangian includes, as special cases,
SYM theories with $\cN=1,2,4$.

In this paper, we will discuss massive deformations of the Lagrangian (\ref{L-undeformed}) which preserve the duality between color and kinematics. We will be particularly interested in theories with trilinear scalar couplings. Upon reduction to four dimensions,
the deformed Lagrangian is
\bea
{\cal L} \!\!\!\! &=& \!\!\!\! - {1 \over 4} (F^{\hat a}_{\mu \nu})^2 + {1 \over 2} (D_\mu \phi^{\hat a I})^2
-{1\over 2 } m^2_{IJ}  \phi^{\hat a I}  \phi^{\hat a J} \!\!
- {g^2 \over 4} f^{\hat a \hat b \hat e} f^{\hat c \hat d \hat e} \phi^{\hat a I} \phi^{\hat b J} \phi^{\hat c I} \phi^{\hat d J} - {g\lambda \over 3!} f^{\hat a \hat b \hat c} F^{IJK} \phi^{\hat a I}  \phi^{\hat b J}  \phi^{\hat c K}  \no \\
&&+ {i \over 2} \bar \psi \cancel{D} \psi - {1 \over 2}  \bar \psi M \psi  + {g \over 2} \phi^{\hat a I} \bar \psi\Gamma^{I} t^{\hat a} \psi \ .
\label{2.5}
\eea
The covariant derivatives are
\bea
D_\mu \phi^{\hat a I} &=& \partial_\mu \phi^{\hat a I} + g f^{\hat a \hat b \hat c} A_\mu^{\hat b} \phi^{\hat c I} \ , \\
D_\mu \psi &=& \partial_\mu \psi  - i g t^{\hat a} A_\mu^{\hat a} \psi \ . \label{L-deformed}
\eea
For the discussion of color/kinematics duality in this section, we will keep general the representation ${R}$ of fermionic fields.
In later sections we will choose it to be the adjoint.
Following the notation in refs. \cite{Chiodaroli:2017ehv, Chiodaroli:2015rdg, Chiodaroli:2015wal}, gauge-theory gauge-group indices are hatted throughout the paper.
Gauge representation indices, fermion global indices, and spacetime spinor indices are not explicitly displayed in eq.~(\ref{L-deformed}). $I,J=4, \ldots , 3+n_S$ are global indices running over the number of scalars in the theory.
We shall choose the scalar mass matrix $m^2_{IJ}$ to be diagonal; moreover, while the fermion mass matrix  $M$ can have off-diagonal entries, we shall assume that its square is also diagonal.

Our conventions
are collected in Appendix \ref{app-conventions}.
We use a mostly-minus metric. The matrix
$\gamma^0$  is Hermitian and we have the relation $(\gamma^\mu)^\dagger = -\gamma^0 \gamma^\mu \gamma^0$ ($\mu=0,1,2,3$). Spinors in the above Lagrangian obey Majorana conditions
\be
\bar \psi = \psi^t { C}_4 \Omega_g C_{D-4} \ ,
\ee
where ${C}_4$ is the four-dimensional charge-conjugation matrix.
$\Omega_g $ and $C_{D-4}$ are matrices acting on the gauge and flavor indices carried by the fermions.\footnote{More concretely, for adjoint fermions $\Omega_g$ is the identity matrix. For fermions in a pseudo-real representation, $\Omega_g$ is a unitary antisymmetric matrix which relates the generators of the gauge-group representation to the generators of the conjugate representation.}
We also note that reality of the Lagrangian \eqref{2.5} requires that
\be
 (\Gamma^I)^\dagger \gamma^0 = -\gamma^0 \Gamma^I \qquad \ \ \text{and} \ \ \qquad M^\dagger \gamma^0= \gamma^0 M \ .
\ee
In the following, it will be convenient to avoid displaying explicitly the flavor/global indices for the fermions; similarly to
spacetime spinor indices, their contraction is realized as matrix multiplication.

Color/kinematics duality of the massless limit of the bosonic part of the Lagrangian was established in \cite{Chiodaroli:2014xia}.
Demanding that it also holds for the  four-fermion, four-scalar and two-fermion-two-scalar amplitudes of
the complete Lagrangian yields constrains on the scalar and fermion mass matrices and the three-index tensor $F^{IJK}$.
We shall derive and solve them in the following sections. A study of the five-point amplitudes reveals no additional constraints.

\subsection{Four-fermion amplitudes}

To write down the four-fermion amplitude it is convenient to introduce the $(4+n_S)$-dimensional Dirac matrices
\be
\Gamma^A = \left\{  \begin{array}{cc} \gamma^\mu \otimes \id & \mu=A<4 \\
~~\gamma_5\otimes \Gamma^I  & I=A \geq 4\end{array}
\right. \ .
\ee
Denoting collectively the spacetime and flavor spinor indices as $a_1, \ldots a_4$, we can write the four-fermion amplitude in a compact form as\footnote{We define $t^{\hat a}_{\hat \imath_1 \hat \imath_2}$ with lower indices as $(\Omega_g t^{\hat a})_{\hat \imath_1 \hat \imath_2}$, i.e.  the representation indices $\hat \imath_1 , \ldots, \hat \imath_4$ for the gauge matrices have been lowered with $\Omega_g$.}
\be
{\cal A}_4\big({1\psi_{ \hat \imath_1 a_1}, 2\psi_{ \hat \imath_2 a_2}, 3\psi_{\hat \imath_3 a_3 }, 4\psi_{\hat \imath_4 a_4 }}) =\sum_A  {i g^2\over s-m^2_{A}} \Big(
(C_D \Gamma^A)_{a_1a_2} (C_D \Gamma_A )_{a_3a_4}\Big)  t^{\hat a}_{\hat \imath_1 \hat \imath_2} t^{\hat a}_{\hat \imath_3 \hat \imath_4} + \text{Perms} \ , \label{4ferm}
\ee
where the matrix $C_D$ is $C_D={ C}_4 C_{D-4}$ and $m^2_{A}$ denotes the mass of the particle exchanged in the $s$ channel. For the four-dimensional components of the index $A$, the exchanged particle is a vector field,  so the mass vanishes;
for $A\ge 4$, this particle is a scalar field so the mass is the relevant entry of the scalar mass matrix in eq.~\eqref{2.5}.

Every term in this amplitude is manifestly gauge invariant so, {\it a priori}, we need not impose any specific correlation between color factor
relations and kinematic factor relations. If however we double-copy eq. (\ref{4ferm}) with vector amplitudes and expect to obtain an amplitude
that is invariant under local supersymmetry, then the numerators should be required to obey relations analogous to those of the
color factors.

For fermions in the adjoint representation the color factors obey the Jacobi  identity
\be
t^{\hat a}_{\hat \imath_1[ \hat \imath_2}  t^{\hat a}_{\hat \imath_3 \hat \imath_4]} = 0 \ .
\label{color_J1}
\ee
Since we will interpret the mass of the exchanged particle as being induced through dimensional reduction and, moreover, it does not appear explicitly in the numerator factors, it is natural that we demand that the numerators obey the same relation as if all exchanged particles were massless. Stripping away the fermion wave functions (since the numerator relation should hold for
all values of momenta), the kinematic relation dual to eq.~\eqref{color_J1} reduces to
\be
(C_D \Gamma^A )_{a_1[a_2} (C_D \Gamma_A )_{a_3 a_4]} = 0 \ ,
\label{fermion_num_rel}
\ee
up to a possible projector enforcing the chirality of fermion wave function.
This condition can be satisfied if the theory is a YM theory with one irreducible spin-$1/2$ fermion in
dimensions\footnote{It also trivially holds in $D=2$ because of over-antisymmetrization.} $D=3,4,6,10$;
it is equivalent to requiring that the supergravity theory obtained from a double copy  has supersymmetry restored in the massless limit
(the gauge theory itself is not supersymmetric if bosons and fermions
have different representations).
Upon dimensional reduction the color-factor relation is decomposed into representations of the unbroken lower-dimensional
gauge group, similarly to the case of vector-field scattering amplitudes. Thus, the numerator relation \eqref{fermion_num_rel}
continues to be the appropriate one for double-copy constructions in which massive fermions combine with massive
$W$ bosons.

Starting from Section \ref{Sec-sol}, we will focus on the case where the range of the indices is $I,J=4,\ldots, 9$, i.e. when the theory is a massive deformation of $\cN=4$ SYM. In this case, the spinors obey the chirality condition
\begin{equation}
 \Gamma_{11} \psi \equiv [\Gamma^{*}_6 \otimes \gamma_5 ] \psi =  \psi \ ,  \label{chirality}
\end{equation}
where $\Gamma_{11}, \gamma_5$ and $\Gamma^{*}_6$ are the chirality matrices in $10,4$ and $6$ dimensions, which we
take  to be Hermitian. We also note that the matrix $M$ needs to obey the condition
\begin{equation}
\{M, \Gamma_{11} \} = 0 \ , \label{Mchirality}
\end{equation}
for the mass term to be consistent with the chirality projection in ten dimensions. These constraints ensure that the number of degrees of freedom is that of $\cN=4$ SYM theory.

\subsection{Two-fermion two-scalar amplitudes}

The two-fermion-two-scalar amplitude given by the Lagrangian \eqref{2.5} is
\bea
{\cal A}_4\big(1\bar \psi^{\hat \imath_1},2 \psi_{\hat \imath_2},3 \phi^{\hat a I},4 \phi^{\hat b J}\big)&=& i g^2 \bar v_2  \Gamma^I {\cancel{p_1}+\cancel{p_4}+M \over (p_1+p_4)^2 + M^2} \Gamma^J u_1 \  (t^{\hat a} t^{\hat b})_{\hat \imath_2}^{\  \hat \imath_1} + (3 \leftrightarrow 4) \no \\
&& \hskip -1.2in +  g^2 \bar v_2 {\cancel{p_3}-\cancel{p_4} \over (p_1+p_2)^2} u_1 \delta^{IJ} \ f^{\hat a \hat b \hat c}(t^{\hat c})_{\hat \imath_2}^{\ \hat \imath_1}  +i g^2 \lambda  F^{IJK} {\bar v_2 \Gamma^K u_1 \over (p_1+p_2)^2 } \  f^{\hat a \hat b \hat c}(t^{\hat c})_{\hat \imath_2}^{\ \hat \imath_1} \ .
\eea
As mentioned previously, we do not display explicitly global indices for the spinor wave-functions and take them contracted
through matrix multiplication. The spinor polarizations obey the massive Dirac equation with a possibly off-diagonal mass matrix
\begin{equation}
(\cancel{p}+M)u = 0 \ , \qquad \bar v (\cancel{p}-M) = 0 \ . \label{Dirac-higher}
\end{equation}
While the spinors $u$ and $v$ should be present in the color/kinematics-duality constraints on the numerators, their
momentum dependence allows us to strip them off; their only remnant is a projector enforcing their chirality properties.
In odd dimensions, the resulting constraints are
\bea
(1) &\qquad& \big\{ \Gamma^I, \Gamma^J \big\} = - 2 \delta^{IJ} \ , \\
(2) &\qquad&\,  \Gamma^I \Gamma^J M + \Gamma^I M \Gamma^J - M \Gamma^J \Gamma^I -\Gamma^J M \Gamma^I + i \lambda F^{IJK} \Gamma^K = 0 \ . \label{master}
\eea
In even dimension, the second equation is modified to include the chirality projector $P_+$:
\be
~~~~~~~~~~~
(2') \qquad  ~~~~\big( \Gamma^I \Gamma^J M + \Gamma^I M \Gamma^J - M \Gamma^J \Gamma^I -\Gamma^J M \Gamma^I + i \lambda F^{IJK} \Gamma^K \big)P_+ = 0 \ . \label{master2}
\ee
A similar version of this relation, in which $M$ also acts on flavor indices,
can be obtained when more than one irreducible spinor is present (as in the theories considered in
ref.~\cite{Chiodaroli:2015wal}).

\subsection{Four-scalar amplitudes}

To find the constraints stemming from color/kinematics duality on the trilinear bosonic
interactions, we analyze the four-scalar amplitudes. The kinematic
numerators determining the scattering amplitude of four massless
scalars following from the Lagrangian \eqref{2.5}, ${\cal A}_4 \big( 1 \phi^I , 2 \phi^{J} , 3 \phi^{K} , 4 \phi^{L} \big)$, are
 \begin{eqnarray}
 n_s &=& \delta^{IJ}\delta^{KL} (t-u) + s (\delta^{IL}\delta^{JK} - \delta^{IK}\delta^{JL}) -  \lambda^2 F^{IJM}F^{KLM} \ , \\
 n_u &=& \delta^{IK}\delta^{JL} (s-t) + u (  \delta^{IJ}\delta^{KL} - \delta^{IL}\delta^{JK}) - \lambda^2 F^{KIM}F^{JLM} \ , \\
 n_t &=& \delta^{IL}\delta^{JK} (u-s) + t ( \delta^{IK}\delta^{JL} - \delta^{IJ}\delta^{KL}) - \lambda^2 F^{JKM} F^{ILM} \ .
 \end{eqnarray}
 Imposing the numerator identity
 \begin{equation}
 n_s + n_u + n_t = 0
 \end{equation}
 results in Jacobi relations that need to be obeyed by the $F^{IJK}$-tensors, which are then identified as the supergravity
 gauge-group structure constants, as explained in ref. \cite{Chiodaroli:2014xia}.

To understand the constraints we impose on scattering amplitudes with massive scalar fields, it is important to recall that,
on the one hand, these amplitudes are double-copied with amplitudes with massive vector fields and, on the other, the resulting
supergravity amplitudes should exhibit standard gauge invariance  from a higher-dimensional perspective \cite{Chiodaroli:2015rdg}.
It is natural
to assign complex representations to massive scalars.
Moreover, when viewed as numerators in a higher-dimensional theory, the scalar amplitudes' numerators should obey standard
Jacobi relations.

The amplitudes with four scalars of identical mass, ${\cal A}_4 \big( 1 \varphi^i , 2 \varphi^{j} , 3 \varphi^{\bar k} , 4 \varphi^{\bar l} \big)$, are given by cubic graphs with numerators\footnote{
Note that other equivalent numerator factors may be used in particular cases. For example, the $s$-channel numerator can be set to zero for massive scalar amplitudes in the ${\cal N}=2^*$ theory and in ${\cal N}=2$ SQCD \cite{Johansson:2014zca,Johansson:2015oia,Johansson:2017bfl}.}
 \begin{eqnarray}
 n_s &=&  (s - m^2_s) (\delta^{i \bar l}\delta^{j \bar k} - \delta^{i \bar k}\delta^{j \bar l}) -  \lambda^2 F^{ ij \bar m}F^{ m \bar k \bar l} \label{n-s-ch}\ , \\
 n_u &=& \delta^{i \bar k}\delta^{j \bar l} (s-t) - u  \delta^{i \bar l}\delta^{j \bar k} + \lambda^2 F^{ai \bar k}F^{aj \bar l} \ , \\
 n_t &=& \delta^{i \bar l}\delta^{j  \bar k} (u-s) + t \delta^{i \bar k}\delta^{j \bar l}  - \lambda^2 F^{a j \bar k} F^{a i \bar l} \ .
 \end{eqnarray}
 The terms independent of the tensor $F$ are due to either vector-field exchange or the quartic-scalar interaction.
 The mass dependence in numerators appears from resolving the quartic scalar vertices into cubic graphs.  Due to
 representation assignment, it is natural to associate a zero mass to the propagators for the $t,u$ channels; the $s$ channel
 could potentially be assigned a nonzero mass $m_s$, as shown.
 This is a consequence of the structure of color and kinematics factors in scattering amplitudes involving massive $W$ bosons which
 double-copy with this amplitude.\footnote{
 Indeed, interpreting the mass of $W$ bosons as momentum in higher dimensions \cite{Chiodaroli:2015rdg} implies that
 higher-dimension color factors obey standard Jacobi relations. Upon dimensional reduction, color factors are decomposed
 following the breaking of the adjoint representation of the higher-dimensional gauge group; while some resulting components
 of the higher-dimensional color factor vanish, they come with a nonvanishing kinematic numerator and contribute to the
 double copy \cite{Bern:2012uf}.}

Requiring the kinematic numerator relation for amplitudes in theories with at least two types of massive scalars implies that
the $F$-tensor and the $s$-channel mass are related as
\begin{equation}
\lambda^2 \left( F^{ ij \bar m}F^{ m \bar k \bar l} - F^{ai \bar k}F^{aj \bar l}  + F^{a j \bar k} F^{a i \bar l}\right) - m^2_s (\delta^{i \bar l}\delta^{j \bar k} - \delta^{i \bar k}\delta^{j \bar l}) = 0  \  .
\label{Jacobi-modify}
\end{equation}
The case of theories with a single massive scalar must be considered separately; in that case the quartic scalar does not contribute
to the $s$-channel because the relevant combination of Kronecker-delta functions, $(\delta^{i \bar l}\delta^{j \bar k} - \delta^{i \bar k}\delta^{j \bar l})$, vanishes by symmetry. Thus, thre is no $s$-channel
mass term and consequently the $F$ tensor should obey a standard Jacobi identity:
\begin{equation}
F^{ ij \bar m}F^{ m \bar k \bar l}  - F^{ai \bar k}F^{aj \bar l}  + F^{a j \bar k} F^{a i \bar l} = 0 \ .
\end{equation}
In the following sections, we will solve the constraints \eqref{master}, \eqref{master2} on $F$
and, if more than one massive scalar are present, we will
determine their masses from eqs.~\eqref{Jacobi-modify}.

\subsection{Solution for general $F^{IJK}$ in $D$ dimensions\label{sec-Dsol}}

Before focusing on the case of $\cN=4$~SYM~theory, which descends from $\cN=1$~SYM~theory in ten dimensions and hence corresponds to a six-dimensional internal space, we may consider the general case with an unconstrained number of internal dimensions.

If we assume that $F^{IJK}$ is given, the general solution of the equation
\be
\Gamma^I \Gamma^J M + \Gamma^I M \Gamma^J - M \Gamma^J \Gamma^I -\Gamma^J M \Gamma^I + i \lambda F^{IJK} \Gamma^K  = 0
\label{SameEqnAgain}
\ee
is the sum of the general solution of the homogeneous ($F^{IJK}=0$) equation and of a particular solution of the
inhomogeneous ($F^{IJK}\neq0$) equation. Thus, defining $X=\{ \Gamma^I,  M  \}$, for the homogeneous solution
we need to solve
\be
 \big[ X, \Gamma^J\, \big]=0\,.
\ee
This implies that $X$ also commutes with all the antisymmetrized products, $\Gamma^{I_1\dots I_n}$ of Dirac matrices.
As these products form a basis in the space of matrices, it follows that $\{ \Gamma^I,  M  \} \propto \mathbb{I}$. In turn this
equation implies that $M$ is a linear combination of $\Gamma$ matrices.  Thus, the general solution of
the homogeneous ($F^{IJK}=0$) part of \eqn{SameEqnAgain} is
\be
M= u_L \Gamma^L\, ,
\label{homogeneoussol}
\ee
where $u_L$ are free parameters of unit mass dimension.

For the particular solutions we may distinguish between two cases: (1) The $F^{IJK}$ are structure constants of a simple Lie algebra, (2) the $F^{IJK}$ have a more general interpretation (e.g. $F^{IJK}$ do not satisfy the Jacobi identity). For the first case, one can argue that $M$ should be linearly related to $F^{IJK}$ and moreover the adjoint Lie algebra indices need to be contracted with gamma matrices for the group symmetry not to be broken. With this in mind, a particular solution to \eqn{SameEqnAgain} is given by
\begin{equation}
M = i {\lambda \over 4! }  F_{IJK} \Gamma^{IJK}  \, .
\label{particularsol}
\end{equation}
To show that \eqn{particularsol} is a solution, one needs only use that $F_{IJK}$ is totally antisymmetric; thus,
eq.~\eqref{particularsol} solves eq.~\eqref{SameEqnAgain} even if $F_{IJK}$ does not obey the Jacobi identity.

Retuning to $F^{IJK}$ being the structure constants of a Lie algebra, it is not difficult to find the corresponding generators.
Defining $T^I\equiv \{ \Gamma^I,  M  \}$, one can show that after anti-commuting \eqn{SameEqnAgain} with $M$ one
obtains
\be
0=\Big\{ \big[\,  \{ \Gamma^I,  M  \} , \Gamma^J\, \big]+ i \lambda F^{IJK} \Gamma^K ,M \Big\} = \big[  T^I, T^J \big]  -\big\{  \big[ M^2,  \Gamma^I  \big], \Gamma^J \big\} + i \lambda F^{IJK} T^K\,.
\label{defLieAlg}
\ee
If, for a moment, we restrict our attention to the case where $M^2$ is proportional to the identity, then the commutator $\big[ M^2,  \Gamma^I  \big]$ vanishes and \eqn{defLieAlg} becomes the defining relation for a Lie algebra. However, such an $M^2\propto \mathbb{I} $ is not the generic situation.  Instead, we can investigate directly the commutation properties of $T^I$ introduced above.
Using their explicit form,
\be
T^I\equiv \{ \Gamma^I,  M  \} =i \frac{\lambda}{4} F^{JIK} \Gamma^J \Gamma^K\, ,
\ee
their commutator is
\be
 \big[  T^I, T^J \big] = \frac{\lambda^2}{4}  \big(F^{ILK} F^{KJM}-F^{IMK} F^{KJL}\big)\Gamma^L \Gamma^M \, \stackrel{\rm Jacobi}{=} \, - i \lambda F^{IJK} T^K\, ,
\ee
where, in the last equality, we used the Jacobi identity for $F^{IJK}$.
Thus, the $T^I$ are Lie-algebra generators precisely
when $F^{IJK}$ are structure constants of a Lie algebra.
From \eqn{defLieAlg}, we can conclude that $\big\{  \big[ M^2,  \Gamma^I  \big], \Gamma^J \big\} = 0$ whenever $F^{IJK}$
comes from a Lie algebra. This relation relies on the Jacobi identity and can be confirmed through direct calculation.
The general solution to \eqn{SameEqnAgain} obtained by combining (\ref{homogeneoussol}) and  (\ref{particularsol}) is
\be
M= u_L \Gamma^L+i {\lambda \over 4! }  F_{IJK} \Gamma^{IJK} \,.
\ee
It is interesting to compute the squared mass matrix. One obtains
\be
M^2 =\frac{1}{2}\{M, M  \}=-\Big(u^I-\frac{1}{2}T^I \Big)^2 - \frac{\lambda^2}{48} F^{IJK}F_{KJI}\,.
\ee
which is in general not proportional to the unit matrix because of the appearance of $T^I$.
Note that in our conventions $M$ is anti-Hermitian, hence $M^2$ is negative definite.

As a further generalization, note that if we do not take the $F_{IJK}$ as given, linearity of eq.~\eqref{SameEqnAgain} allows us to superpose several particular solutions for different structure constants and couplings, through the replacement $\lambda F_{IJK} \rightarrow \lambda F_{IJK} + \lambda' F'_{IJK}+ \ldots$ in all of the above formulas.
In case the different structure constants have non-overlapping indices, this operation will result in a gauged supergravity with a product gauge group.
If the indices in the structure constants  overlap, we can no longer interpret them as belonging to the adjoint of a given gauge group. Ultimately, whether the $F$-tensors obey a conventional or modified Jacobi identity is the result of imposing color/kinematics duality on four-scalar amplitudes in the gauge theory entering the double-copy construction.

\section{Massive deformations of $\cN=4$ SYM theory \label{Sec-sol}}
\renewcommand{\theequation}{3.\arabic{equation}}
\setcounter{equation}{0}

To focus on double-copy constructions of gaugings of $\cN=8$ supergravity with Minkowski vacua (which also posses some unbroken global symmetry), we discuss in detail
solutions to the constraint (\ref{master}) which reduce to $\cN=4$ SYM theory in the massless limit. To this end, we will take
fermions to also transform in  the adjoint representation of the gauge group and we will set the number of scalar fields $n_S=6$.
We will start from a massive deformation of $\cN=4$ SYM which solves eq.~(\ref{master}) in dimension higher than four and
use a combination of Higgsing and orbifolding to construct a four-dimensional theory which still obeys color/kinematics duality.
Hence, it will make sense to collect the solutions of our constraint into two groups according to whether or not they admit an uplift
to  dimension higher than four.

\subsection{Solutions that uplift to $D>4$}

Solutions which admit uplift to higher dimensions, organized following their unbroken symmetry and mass spectra, are:
\begin{enumerate}

	\item[i.] {\bf  $SO(5)$}. We take
	\be M =  u   \Gamma^{9} \ , \qquad  F^{IJK} \equiv 0  \ . \ee
	The fermion mass term in this case is the one obtained by giving a vacuum expectation value to $\phi^{9}$.
	Only one fermionic mass is present in the spectrum. The solution can be uplifted up to $9D$ and can also be
	obtained from the spontaneously-broken theory of ref. \cite{Chiodaroli:2015rdg} by orbifolding away
	bosonic fields in massive vector multiplets and fermionic fields in massless vector multiplets.
	
	\item[ii.] {\bf  $SU(2) \times SU(2)$: the interesting solution}. We take
	\be M = i {\lambda \over 4}  \Gamma^{789}  \ , \qquad  F^{789} = 1    \ . \ee
	There is a single  mass in the spectrum which is given by\footnote{In our convention the physical masses are given by the eigenvalues of $-M^2$ ($M$ is negative-definite).} $m_1={\lambda / 4}$. This solution is the analog of the $D$-dimensional solution presented in Section \ref{sec-Dsol}  and can be uplifted up to seven dimensions. Note that only one of the two $SU(2)$ factors is reflected by the trilinear scalar couplings.
		
	\item[iii.] {\bf $SU(2)_R$: the $\cN=2^*$ theory}. To look for solutions preserving some supersymmetry, we employ a
	complex basis for the Dirac matrices, splitting the $SO(6)$ index $I$ as $I=(4,5,1,\bar 1, 2 , \bar 2)$. This decompostion
	makes the
	$SU(2)\subset SU(4)$ subgroup manifest which will be the surviving R-symmetry. A natural Ansatz preserving
	$SU(2)$ symmetry is
	\begin{equation}
	M =  {\lambda \over 4} \Gamma^{5 1 \bar 1} + {\lambda \over 4} \Gamma^{5 2 \bar 2} \ , \qquad
	F^{5 i \bar \jmath} = i  \delta^{i \bar \jmath}  \ .
	\end{equation}
	It is easy to see that the square of the mass matrix is proportional to a half-rank projector,
	\begin{equation}
	M^2 = -  {\lambda^2\over 8}\Big( 1 + \Gamma^{1 \bar 1 2 \bar 2} \Big) \ ,
	\end{equation}
	suggesting that this choice corresponds to the $\cN=2^*$ theory. Since in this case supersymmetry requires that
	two scalar fields are massive, the relevant numerator identity is (\ref{Jacobi-modify}). It fixes the mass for the
	$s$-channel exchange to be
	\begin{equation}
	F^{5 1 \bar 1} F^{5 2 \bar 2 }  - m_s^2 = 0 \quad \rightarrow  \quad  m_s= \lambda  \ . \label{numN2s}
	\end{equation}
	Hence, the two complex scalars have mass $m_1=\lambda/2$.\footnote{The mass of the $s$-channel ($\lambda$) is twice the mass of individual external states  ($\lambda/2$) this is a consequence of mass conservation in the vertices of the theory, which in turn can be related to the flow of compact momenta.} This theory
	can be uplifted to five dimensions.
	
	\item[iv.] {\bf $SO(2)\times SO(2)$: hybrid solutions}. Since the constraint (\ref{master}) is linear in $M$, further
	solutions can be obtained by linearly superposing solutions (i) and (ii) or solutions (i) and (iii).\footnote{Solutions
	(ii) and (iii) can also be superposed, but the result has no surviving symmetry and we do not consider such cases here.}
	For example, a three-parameter family of solutions is given by
	\be
	M =  i {\lambda \over 4}  \Gamma^{789} + u_1 \Gamma^{9} +
	u_2 \Gamma^{6}  ,  \quad F^{789} = 1 \ .
	\ee
	There are two distinct fermionic masses:
	\begin{equation}
	m^2_{1,2} = (\lambda/4 \pm u_1)^2 + (u_2)^2 \ .
	\end{equation}
	We note that, in this case, the gauge symmetry of a supergravity obtained through the double copy is
	spontaneously broken. This solution can be uplifted up
	to $6D$ (or $7D$, if $u_2=0$).
	\end{enumerate}
We stress that the list above is not exhaustive. Rather, it reflects our choice of focusing on gaugings of $\cN=8$ supergravity which possess some residual global symmetry and a non-Abelian unbroken gauge group.

\subsection{Solutions in $D=4$}

While we shall not discuss constructions that are indigenous to four dimensions, we include here for completeness
solutions to eq.~\eqref{master} that cannot be uplifted to higher dimensions:
\begin{enumerate}
	\item[v.] {\bf  $SU(2)\times SU(2)$}. We take
	\be M = i {\lambda_1 \over 4}  \Gamma^{789} + i {\lambda_2 \over 4}  \Gamma^{456} \ , \qquad \lambda F^{789} = \lambda_1, \qquad \lambda F^{456} = \lambda_2   \ . \ee
	The mass is given by $m^2=\lambda_1^2/16 +\lambda_2^2/16$.
	
	\item[vi.] {\bf $SO(2)\times SO(2)$}. A four-parameter family of solutions is obtained with
	\be M =  i {\lambda_1 \over 4}  \Gamma^{789} +  i {\lambda_2 \over 4} \Gamma^{456}  +  u_1 \Gamma^{6} +
	u_2 \Gamma^{9}  ,  \quad \lambda F^{789} = \lambda_1, \quad \lambda F^{456} = \lambda_2 \ .  \ee
	This solution is in a sense a superposition from solutions (i) and (v). There are four distinct fermionic masses given by
	\begin{equation}
	m^2_{1,2,3,4} = (\lambda_1/4 \pm u_1)^2 + (\lambda_2/4 \pm  u_2)^2 \ .
	\end{equation}.
	
\end{enumerate}

\subsection{Relation between gauge-theory trilinear couplings and supergravity gauge group}

Looking ahead to the supergravity theories produced by the double-copy construction with the ingredients presented
in this section, it is important to point out that the symmetries used in this section to classify the massive deformations
do not necessarily become gauged in the resulting supergravity theory. Rather, the non-Abelian part of the supergravity
gauge symmetry can be related to the symmetries of the trilinear scalar couplings.

In particular:

\begin{itemize}
	\item If three massless adjoint scalars enter a trilinear coupling in one of the gauge theories, the corresponding $F^{IJK}$-tensor will give (part of) the structure constants of the supergravity unbroken gauge group.
	\item If two massive (matter) and one massless (adjoint) scalars enter a trilinear coupling in one of the gauge theories, the supergravity theory will have an unbroken $U(1)$ factor under which the supergravity fields constructed out of the massive scalars will be charged.
	\item If a spontaneously-broken gauge theory enters the construction, a $U(1)$ factor is also produced since these theories generically have couplings between the massless scalar acquiring the VEV and pairs of massive scalars.
	\item In general, supergravity $U(1)$ photons will be linear combinations of massless vectors with different double-copy origins. The correct linear combination is often obtained by the requirement that  vector fields which remain massless are not charged under the $U(1)$ symmetry.
\end{itemize}
For example, the deformation (ii) is characterized by its $SU(2) \times SU(2)$ global symmetry. Its trilinear couplings, however, correspond to a single $SU(2)$ factor in the supergravity gauge group. \\

\section{Gaugings of $\cN=8$ supergravity with Minkowski vacua \label{Sec-DC}}
\renewcommand{\theequation}{4.\arabic{equation}}
\setcounter{equation}{0}

In this sections, we use the theories identified in the previous sections to formulate double-copy constructions for amplitudes in three distinct gaugings of $\cN=8$ supergravity. Our construction follows three-steps:
\begin{enumerate}
	\item We start from a massive deformation of $\cN=4$ SYM which satisfies the constraint (\ref{master}) in dimension
	higher than four.
	\item We take the theory on the Coulomb branch by assigning compact momenta to some of the fields, as explained in ref. \cite{Chiodaroli:2015rdg}. This step generates various matter fields which are, in general, not in the adjoint representation.
	\item We use an orbifold projection \cite{Chiodaroli:2013upa} to truncate away some of the fields such that,
	 through double copy, the remaining ones reproduce in the massless limit the field content of ungauged $\cN=8$ supergravity.
\end{enumerate}
The last two operations preserve color/kinematics duality,
and hence yield a lower-dimensional theory which can be used in the double-copy method.

Indeed, assigning compact (higher-dimensional) momenta to some of the fields breaks the gauge group $G$ into a (not necessarily semisimple) subgroup, with respect to which massive fields transform in various (not necessarily irreducible) matter representations. In most cases, in the following sections we will take $G=SU(N)$ for some arbitrary $N=N_1+ \ldots + N_k$. The unbroken gauge group will be $SU(N_1) \otimes \ldots \otimes SU(N_k) \otimes U(1)^{k-1}$ and massive fields will transform in various bifundamental representations. Since masses  in the lower-dimensional theory are given by the compact momenta,
color/kinematics will be inherited from the higher-dimensional unbroken theory.

Next, we identify a discrete subgroup $\Gamma \subset G \times SU(4)$ which leaves invariant the compact momenta
(or, from the perspective of the lower-dimensional theory, the VEVs of scalar fields) and project out at the level of the amplitude
all asymptotic states which are not invariant under the transformations
\begin{equation}
\Phi \rightarrow R_i g_i^\dagger \Phi g_i \ , \qquad \forall \ (R_i,g_i) \ \in \Gamma \ .
\end{equation}
Here $\Phi$ denotes a generic gauge-theory field.
Since both $SU(4)$ and $G$ are symmetries of the Lagrangian and $\Gamma$ is preserved by the choice of compact momenta,
the orbifold realizes a consistent truncation. Numerator relations in the truncated theory are inherited from those of the parent
theory. The orbifold construction can be easily implemented by introducing the projector
\begin{equation}
{\cal P}_\Gamma \Phi = {1 \over |\Gamma|} \sum_{(R_i,g_i) \in \Gamma} R_i g_i^\dagger \Phi g_i   \ ,
\end{equation}
where $|\Gamma|$ is the order of the orbifold group.
We note that, for theories containing fermions, we need to impose an additional consistency requirement which guarantees
that CPT-conjugate pairs of fermionic states survive the projection. This translates into
\begin{equation}
{\cal P}_\Gamma \Lambda_g C_6^{-1} = \Lambda_g^* C_6^{-1} {\cal P}_\Gamma^* \ , \label{CPT-cond}
\end{equation}
where $\Lambda_g$ is a matrix that acts on gauge indices and maps each representation into its conjugate.

The final product of the above construction is a theory which contains massless fields in the adjoint representation of the
unbroken gauge group\footnote{Since the unbroken gauge group is a product group, the adjoint representation is understood as
the sum of adjoint representations of all the factors.} together with various matter representations. Combining the numerators of pairs of gauge theories which obey color/kinematics duality leads, through the double-copy relation
\begin{equation}
{\cal M}^{(L)}_{n} = i^{L-1}\;\!\Big(\frac{\kappa}{4}\Big)^{n-2+2L} \sum_{i \in \text{cubic}}\, \int \frac{d^{LD}\ell}{(2\pi)^{LD}} \frac{1}{S_i} \frac{n_i \tilde{n}_i}{D_i}
\label{DCformula} \ ,
\end{equation}
to gravitational amplitudes  (i.e. amplitudes invariant under linearized diffeomorphisms).
At the level of the spectrum, pairs of gauge-theory states correspond to a supergravity state only when they transform
in conjugate gauge-group representations and they have the same mass. Consequently,
in the above formula, we  combine pairs of numerators only when all lines in the corresponding graphs carry conjugate
representations and have propagators with the same mass squared.

\subsection{Consistency requirements}

As a consequence of the particular kind of double-copy procedure we are adopting, several consistency requirements need to be obeyed by the gauge theories entering the construction, which include:

\begin{figure}
\begin{center}
\includegraphics[width=0.8\textwidth]{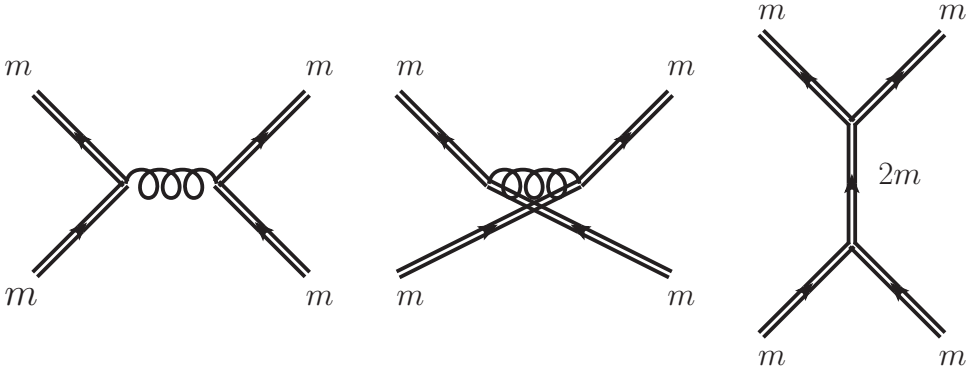}
\caption{Diagrams generically arising when a theory is taken on the Coulomb branch through assigning compact momenta. In some cases, the third diagram has vanishing color factor but is formally assigned a nonzero numerator factor. Mass flow follows the direction of the arrow and is conserved at each vertex.}
\end{center}	
\end{figure}

\begin{itemize}
	\item {\bf Mass matching} of spectra between the two gauge theories. Namely, the double copy requires that only gauge-theory numerators corresponding to the same set of propagators can be combined in the
	double-copy formula.\footnote{It should be noted that there exist double copies in which this requirement is relaxed, e.g. the double-copy construction for conformal supergravities.} Note that because of this requirement, we will take  the mass matrices squared to be diagonal.
	Enforcing this requirement relates the VEV parameters on the two sides of the double copy.
	\item {\bf No massive gravitons}. Double copies which involve two massive $W$ bosons combining to give a massive graviton should be excluded.
	This amounts to requiring that $W$ bosons from the two gauge theories cannot be assigned to conjugate representations (which in turn imposes nontrivial constraints on the choice of orbifold projectors; such constraints are implemented on a case-by-case basis).
	\item {\bf No Kaluza-Klein towers}.
	As we pointed out previously, there exist amplitudes in a Coulomb-branch theory that have certain vanishing color factors
	but non-zero kinematic numerator factors. This situation arises because Coulomb-branch amplitudes are obtained from
	higher-dimensional massless theories by (1) assigning compact momenta to some of the higher-dimensional massless
	fields, resulting in lower-dimensional masses and (2) modifying the color factors to reflect the breaking of a larger gauge
	group.
	These steps do not alter the higher-dimensional numerator relations, which generically will be three-term relations mirroring the
	color factors of that theory.
	However, the second step sets some color factors to zero,\footnote{This is because only certain combinations of
	 representations of the unbroken gauge group can label the structure constants of the original group.} thereby turning
	 three-term color-factor relations into two-term relations and associating formally a vanishing color factor to a nonzero
	 numerator factor.
	If two sets of such numerators  are combined in the double-copy formula,
	the resulting amplitude exhibits poles  that correspond to additional massive states which are not present in the expected
	lower-dimensional spectrum.
	 These poles correspond  to physical asymptotic states which need to be added to the theory, thereby reconstructing the full
	 Kaluza-Klein tower of massive states. Since we will be interested only in theories with a finite number of fields, we will need
	 to make sure that, whenever one color factor vanishes in one theory, the corresponding kinematic numerator vanishes in
	  the other theory (and vice versa).
	
	\item {\bf States of $\cN=8$ multiplet in the massless limit}. Since orbifolding generates states in various matter representations, we will need to make sure that the spectrum of ungauged $\cN=8$ supergravity is recovered in the massless limit. In some cases, this will force us to combine different irreducible matter representation into a single reducible representation, to avoid multiple copies of the corresponding supergravity states after the double-copy.\footnote{This is analogous in spirit with treating the adjoint representation of a product gauge group as a single (reducible) representation rather than decomposing it into its irreducible components. This interpretation eliminates the possibility of multiple gravitons.}
	Concretely, after Higgsing and orbifolding,  a gauge-theory Lagrangian can be written explicitly in terms of the
	representation matrices $t_{R_1}$ and $t_{R_2}$. Combining these representations implies that the Lagrangian be
	rewritten only in terms of $t_{R_1 \oplus R_2}$.
        In turn, this requires that parameters of the Lagrangian related to the VEVs of  the scalar fields (or compact momenta) be
        related in specific ways.
	To illustrate this point, let us consider the couplings between massive and massless scalars in the spontaneously-broken Lagrangian. They have the form
	\cite{Chiodaroli:2015rdg}
	\begin{equation}
	  - 2 \sum_i m_i h^\ha \overline \varphi t^\ha_{R_i} \varphi \ ,
	\end{equation}
	where here $i$ labels the matter representations generated by spontaneous symmetry breaking, $h$ is the Higgs field
	and $m_i$ is the mass (with sign) corresponding to the $i$-th representation. These couplings  can be rewritten  in terms of the representation matrices of a direct
	sum of two representations, say $t_{R_1 \oplus R_2}$, only if the corresponding $m_i$ are equal.
\end{itemize}

\begin{table}[th]
	\begin{center}
		\small
		\begin{tabular}[h]{c|c|c|c}
			& \bf Orbifold of Coulomb  & \bf Orbifold of    & \bf Orbifold     \\
			& \bf branch of $\cN=4$ SYM     &  \bf $\cN=2^*$ theory &  \bf of  YM theory  \\
			&   (no residual supersymmetry)		& 				& \bf  + massive ferms (ii)     	    \\
			\hline
			&&& \\
			\bf  Coulomb branch & CSS gauging  &  CSS gauging	 & gauged  $\cN=8$   \\
			\bf of $\cN=4$ SYM  &  with $\cN=4$ 		 &  with $\cN=6$     &  supergravity  \\
			\bf         &   	 residual susy				 		&   residual susy 	& with $SU(2)  \times U(1)$   \\
			&            			 		&					&  gauge group and       \\
			& 						 		&					&  $\cN=4$ residual susy                   \\
			\hline
	\end{tabular}
		
		\bigskip
	\end{center}
	\caption{Double-copy constructions of gravitational theories from the theories discussed in Section \ref{Sec-sol}, to be discussed in the next section.\label{tab-DC}}
\end{table}
After all requirements are imposed, we have two possible families of constructions. The first has the schematic form:
\begin{eqnarray}
\left(\begin{array}{c} \text{spontaneously-broken} \\ \text{gauging } \end{array} \right) = \Big( \text{Higgs-YM theory} \Big) \otimes \Big(\cancel{\rm S}\text{YM theory}  \Big) .
\end{eqnarray}

In Table \ref{tab-DC}, we list some possible double copies of this class, which will be discussed in detail in Section \ref{Sec-examples}.
We emphasize that each entry corresponds to a different orbifold projection of the higher-dimensional theories on the Coulomb-branch. The consistency requirements discussed in this section are imposed on a case-by-case basis.

Another class of constructions involves two explicitly-broken gauge theories. The resulting supergravity theory
will not possess spontaneously-broken supersymmetry, and can be thought of as a gauging of $\cN=8$ supergravity
with additional explicit supersymmetry-breaking terms. The schematic form will be:
\begin{eqnarray}
\left(\begin{array}{c} \text{explicitly-broken} \\ \text{gauging } \end{array} \right) = \Big( \cancel{\rm S}\text{YM theory}  \Big) \otimes \Big(\cancel{\rm S}\text{YM theory}  \Big) .
\end{eqnarray}
While these theories are not the main focus of this paper, we will also discuss some examples of this class of theories  in the following section.

\section{Examples \label{Sec-examples}}

In this section, we present various examples of double-copy constructions for gaugings of $\cN=8$ supergravity. We discuss separately the case of spontaneous and explicit supersymmetry breaking and work out spectra and unbroken gauge groups for the theories under consideration. For the theory with $SU(2)\times U(1)$ unbroken gauge group and spontaneously-broken supersymmetry, we also work out selected three-point amplitudes.

\subsection{Spontaneously-broken gaugings}

\subsubsection{CSS gauging with $\cN=6$ unbroken supersymmetry}

Our simplest example involves, as the first (left) gauge-theory factor, an orbifold of the $\cN=2^*$ theory. The gauge group is taken to be $SU(2N)$. The $\cN=2^*$ theory has a $U(1)$ symmetry acting as a complex phase on one half-hypermultiplet and as the conjugate phase on the other. Denoting with $R(\theta)$ the corresponding group element, we orbifold the first theory as
\begin{equation}
\Phi \rightarrow R\Big({2 \pi \over 3}\Big) g^\dagger \Phi g \ , \qquad
g_{\text{left}} = \text{diag}\big( I_{N}, e^{{4\pi \over 3}i} I_{N} \big) \ , \end{equation}
where $\Phi$ denotes a generic field of the theory.
The net result  of this operation is a theory with $SU(N)\times SU(N) \times U(1)$ unbroken gauge group and matter fields in
its bifundamental representation. Writing an $SU(2N)$ group element as a $2\times 2$ block matrix with $N\times N$ entries,
the fields of the theory populate it as
\begin{equation}
\left( \begin{array}{cc} {\cal V}^{(0)}_{\cN=2}  & { \Phi}^+_{\cN=2} \\
{ \Phi}^-_{\cN=2} & {\cal V}^{(0)}_{\cN=2}  \end{array} \right) \ ,
\end{equation}
where ${\Phi}^\pm_{\cN=2}$ are the two half-hypermultiplets which transform in the bifundamental representation $R$ of $SU(N)\times SU(N)\times U(1)$ and its conjugate $\bar R$.
For the second (right) gauge theory we take $\cN=4$ SYM on the Coulomb branch, with VEV given to one of the scalars as
\begin{displaymath}
\langle \phi^4 \rangle = {1 \over g} \text{diag} \big( u I_N, -u I_N \big)   \ .
\end{displaymath}
In this simple example, we do not need to combine any representations and $u$ is the only free parameter.
Mass matching implies that the hypermultiplet mass be equal to $u$. The resulting spectrum is given in Table \ref{Tab-CSS6} and corresponds to the spectrum of CSS gauging with $\cN=6$ supersymmetry.\footnote{We should note that CSS gaugings can be interpreted as gauging  noncompact semidirect product groups of the form $U(1) \ltimes T^m$ where $m\leq 24$ \cite{Andrianopoli:2002mf}. In this case $m=12$ and the 12 massive vector fields correspond to the 12  non-compact translation generators. The vacuum of the theory is invariant  only under the compact $U(1)$ subgroup. } The unbroken gauge group is $U(1)$, as it can be inferred from the trilinear scalar coupling between $\phi^4$ and two massive scalars in the spontaneously-broken gauge-theory factor.

\begin{table}[tb]
	\begin{center}
		\begin{tabular}{ccccc}
			Rep. & R & L & Sugra fields & mass${}^2$  \\	
			\hline \\
			$G$ & ${\cal V}^{(0)}_{\cN=2}$ & ${\cal V}^{(0)}_{\cN=4}$ & ${\cal H}_{\cN=6}$   & $0$  \\
			$R$ & ${\Phi}^+_{\cN=2}$ & ${\cal V}^{(m)}_{\cN=4}$ & ${\Psi}^{(m)}_{\cN=6}$ & $u^2$  \\
			$\bar R$ & ${\Phi}^-_{\cN=2}$ & ${\cal V}^{(m)}_{\cN=4}$ & ${\Psi}^{(m)}_{\cN=6}$ & $u^2$ \\
		\end{tabular}
		\caption{Fields and mass spectra for the CSS gauging of $\cN=8$ supergravity with $\cN=6$ residual supersymmetry. ${\cal V}^{(0)}_{\cN=N}$ and ${\cal V}^{(m)}_{\cN=N}$ denote massless and massive vector multiplets, while ${\cal H}_{\cN=N}$ and ${\Psi}^{(m)}_{\cN=N}$ denote  graviton and massive gravitini multiplets, respectively. ${\Phi}^\pm_{\cN=2}$ denote the two half-hypermultiplets.} \label{Tab-CSS6}
	\end{center}
\end{table}	

It is instructive to also consider  a different realization of the same gauging which arises by starting from a $SO(N+2)$ gauge group. The left gauge theory is obtained orbifolding by
\begin{equation}
\Phi \rightarrow R\big( \pi \big) g^\dagger \Phi g \ , \qquad
g_{\text{left}} = \tilde R (\pi)  \ , \end{equation}
where $\tilde R(\theta)$ is a freely-chosen rotation matrix in $SO(N+2)$ and $R(\theta)$ is the same $U(1)$ element as before. The result of the orbifold projection is a theory with an $\cN=2$ vector multiplet in the adjoint of $SO(N)\times SO(2)$ and a hypermultiplet in the bifundamental representation $(N,2)$. The right theory is $\cN=4$ SYM on the Coulomb branch with a VEV realizing the breaking pattern $SO(N+2) \rightarrow SO(N) \times SO(2)$,
\begin{displaymath}
\langle \phi^4 \rangle = {u \over g} \left( \begin{array}{cccc} 0 & 1 & \cdots & 0 \\
-1 & 0 & \cdots & 0 \\
\cdots & \cdots & & \cdots \\
0 & 0 & \cdots & 0 \end{array}  \right) \ .
\end{displaymath}
We have chosen without any loss of generality  the matrix $\tilde R(\theta)$ to act nontrivially on the first two directions.
The main difference of this variant of the construction is that both the hypermultiplet and the massive $\cN=4$ vector multiplet are in real representations. The spectrum is equivalent to the one given in Table \ref{Tab-CSS6}, but now the bifundamental representation and its conjugate are not treated as distinct and the orbifold action does not separate the two halves of the hypermultiplet.

\subsubsection{CSS gauging with $\cN=4$ unbroken supersymmetry}

The CSS gauging with $\cN=4$ residual supersymmetry is the closest analog to the construction given in ref. \cite{Chiodaroli:2017ngp} and is obtained by starting from a $SO(N+2)$ gauge group and orbifolding by
\begin{equation}
\Phi \rightarrow (-1)^F g^\dagger \Phi g \ , \qquad
g_{\text{left}} = \tilde R (\pi) \ , \end{equation}
where $\tilde R(\theta)$ is the same matrix as in the previous subsection and $F$ is the fermion number. The orbifold projection breaks $SO(N+2)\rightarrow SO(N) \times SO(2)$ with the surviving fields organized as
\begin{equation}
\left( \begin{array}{cc} A_\mu , \phi^i  & \psi^r \\
\psi^r & A_\mu , \phi^i  \end{array} \right) \ ,
\end{equation}
with $i=1,\ldots , 6$ and $r=1, \ldots , 4$. Hence, in this case the orbifold construction breaks supersymmetry completely. We double-copy this theory with $\cN=4$ SYM on the Coulomb branch with VEV as in the previous subsection,
\begin{displaymath}
\langle \phi^4 \rangle =  {u \over g } \left( \begin{array}{cccc} 0 & 1 & \cdots & 0 \\
-1 & 0 & \cdots & 0 \\
\cdots & \cdots & & \cdots \\
0 & 0 & \cdots & 0 \end{array}  \right)   \ .
\end{displaymath}
The spectrum is listed in Table \ref{Tab-CSS4} and the gauge group is $U(1)$, as before.

\begin{table}[tb]
	\begin{center}
		\begin{tabular}{ccccc}
			Rep. & R & L & Sugra fields & mass${}^2$  \\	
			\hline \\
			$G$ & $A_\mu , \phi^i$ & ${\cal V}^{(0)}_{\cN=4}$ & ${\cal H}_{\cN=4}\oplus 6 {\cal V}^{(0)}_{\cN=4}$   & $0$  \\
			$R$ & $\psi^r$ & ${\cal V}^{(m)}_{\cN=4}$ & $4{\Psi}^{(m)}_{\cN=4}$ & $u^2$  \\
		\end{tabular}
		\caption{Fields and mass spectra for the CSS gauging of $\cN=8$ supergravity with $\cN=4$ residual supersymmetry. } \label{Tab-CSS4}
	\end{center}
\end{table}	

\subsubsection{$SU(2) \times U(1)$ gauging with $\cN=4$ unbroken supersymmetry \label{ex-N4}}
This is our main non-Abelian example.
We start from  a copy of the theory (ii) from Section \ref{Sec-sol} in seven dimensions and  take the gauge group to be $SU(3N)$.  We orbifold the theory by projecting  out states which are not invariant under the transformation
\begin{equation}
\psi \rightarrow e^{{2 \pi \over 5}\Gamma_{56}} g^\dagger \psi g \ , \qquad  \phi \rightarrow R_{56}\Big( {4 \pi \over 5}\Big)  g^\dagger \phi g, \ \  \qquad g = \text{diag}\big( I_N, e^{i{2 \pi \over 5}} I_N, e^{i{4 \pi \over 5}} I_N \big) \ ,
\end{equation}
where $R_{56}(\theta)$ is the matrix generating rotations on the $5$-$6$ plane. To make sure that this transformation is
a symmetry of the Lagrangian, we take the scalar mass-matrix to be
\begin{displaymath}
m_{55}= m = m_{66} \ , \qquad {m_{IJ}=0} \ \ \text{ otherwise} \ .
\end{displaymath}
After the projection, the fields of the theory are organized as follows:
\begin{equation}
\left( \begin{array}{ccc} A_\mu , \phi^i & \psi^r & \phi^+ \\
\tilde \psi^{r'} & A_\mu, \phi^i & \psi^r \\
\phi^- & \tilde \psi^{r'} & A_\mu, \phi^i   \end{array} \right) \ ,
\end{equation}
where $i=4,7,8,9$, $r=1,2$, $r'=3,4$, and $\phi^\pm= \phi^5\pm i \phi^6$. In order to end up with a number of states reproducing the spectrum of $\cN=8$ supergravity, we need to combine the representations $(N,\bar N, 1)$ with $(1, N,\bar N)$ and the representation $ (\bar N, N, 1)$ with $(1, \bar N, N)$ into (reducible) representations which are denoted as $R_1, \bar R_1$.
The representations are organized as follows:
\begin{equation}
\left( \begin{array}{ccc} G  & R_1 & R_2 \\
\bar R_1 & G & R_1 \\
\bar R_2 & \bar R_1 & G  \end{array} \right) \ .
\end{equation}
We then want to double-copy this theory with $\cN=4$ SYM on the Coulomb branch.
This theory is obtained by taking a VEV of the form
\begin{displaymath}
\langle \phi^4 \rangle = {1 \over g} \text{diag} \big( u_1 I_N, u_2 I_N, u_3 I_N \big)  \ , \qquad u_1+u_2+u_3=0 \ .
\end{displaymath}
Since the representations which have been combined need to have the same VEV parameter (or, alternatively, the same
mass), we get the condition
\begin{displaymath}
u_1 - u_2 = u_2 - u_3 \quad \rightarrow \quad u_2 = {u_1+u_3\over 2} = 0 \ .
\end{displaymath}
In addition, by matching the mass spectra of the two theories we find that
\begin{eqnarray}
M^2 = -{\lambda^2 \over 16} = - u_1^2 \ ,  &\qquad & m^2  = 4 u_1^2 \ .
\end{eqnarray}
We list the fields from the double copy with their respective mass spectra in Table \ref{Tab-N4}.

\begin{table}[tb]
	\begin{center}
		\begin{tabular}{ccccc}
			Rep. & R & L & Sugra fields & mass${}^2$  \\	
			\hline \\
			$G$ & $A_\mu \oplus \phi^i$ & ${\cal V}^{(0)}_{\cN=4}$ & ${\cal H}_{\cN=4} \oplus 4 {\cal V}^{(0)}_{\cN=4}$  & $0$  \\
			$R_1$ & $\psi^r$ & ${\cal V}^{(m)}_{\cN=4}$ & $2{\Psi}^{(m)}_{\cN=4}$ & $u_1^2$  \\
			$\bar R_1$ & $\tilde \psi^{r'}$ & ${\cal V}^{(m)}_{\cN=4}$ & $2{\Psi}^{(m)}_{\cN=4}$ & $u_1^2$   \\
			$R_2$ & $\phi^+$ & ${\cal V}^{(m)}_{\cN=4}$& ${\cal V}^{(m)}_{\cN=4}$ & $4u_1^2$  \\
			$\bar R_2$ & $\phi^-$ & ${\cal V}^{(m)}_{\cN=4}$ & ${\cal V}^{(m)}_{\cN=4}$ & $4u_1^2$  \\
		\end{tabular}
\caption{Fields and mass spectra for the gauging of $\cN=8$ supergravity with $\cN=4$ residual supersymmetry. The spectrum and unbroken gauge symmetry in the ground state correspond to the $CSO^*(4,4)$ with noncompact gauge group $SO^*(4)$ and maximal compact subgroup $SU(2)\times U(1)$ in the notation of \cite{Catino:2013ppa}.} \label{Tab-N4}
	\end{center}
\end{table}	
The vacuum of this theory has an unbroken $SU(2)\times U(1)$ gauge group  and preserves $\cN=4$ supersymmetry. The $SU(2)$ factor in the gauge group originates from the trilinear coupling between three massless scalars of the theory (ii) from Section \ref{Sec-sol}. This gauging has been previously obtained in ref. \cite{Catino:2013ppa} using the embedding tensor formalism. It was labelled as a $CSO^*(4,4)$ gauging with noncompact gauge group $SO^*(4)$ whose maximal compact subgroup is $SU(2)\times U(1)$.\footnote{Here we should point out that a subtle issue regarding the uniqueness of the identification of the gauged supergravity theory obtained by the double-copy methods. It was shown in \cite{Malek:2017cle} that there exist two inequivalent gaugings of maximal supergravity theory with the same gauge group $[SO^*(4)\times SO^*(4)] \circledS T^{16}$  and identical spectra which have different uplifts to ten dimensions. These two gaugings are related by some outer automorphism. Our identification is modulo such possible ambiguities.} The two massive noncompact gauge fields of $SO^*(4)$ belongs to two massive $\cN=4$ BPS multiplets.

To gain additional information on this gauging, we study amplitudes between two massive and one massless vectors. We choose specific polarizations for the external states and write the massive spinor polarizations as
\begin{eqnarray} &&
u_- =  \left( \begin{array}{c} - M U_- {|q ]\over [ i^\perp q ]} \\ U_- |i^\perp \rangle \end{array} \right) , \qquad
 \bar v_+ =  \left( \begin{array}{cc} U_-^\dagger [i^\perp | & - U_-^\dagger  {\langle q | \over \langle i^\perp q \rangle} M \end{array} \right)  \ .
\end{eqnarray}
Here $U_\pm$ are four component Weyl spinors with positive/negative $6D$ chirality (see Appendix \ref{app-conventions} for details).
The relevant double-copy amplitudes are\footnote{We note that an extra factor of $1/2$ comes from the normalization of the gauge-group generators in the double-copy formula.}
\begin{eqnarray}
A\big ( \bar \psi_+ \psi_- \phi^i \big ) \otimes
A\big ( \bar \psi_+ \psi_- A_+ \big ) & \!\!\!\!\!\! =& \!\!\!\!\!\! {-i \kappa \over 2 \sqrt{2}} \Omega \! \left( \!\!
(U^\dagger_{1-} M \Gamma^i  U_{2-}) (V_{1-}^\dagger  V_{2-}) -
{k_1 \cdot q \over k_2 \cdot q} ( U^\dagger_{1-}  \Gamma^i M  U_{2-}) (V_{1-}^\dagger  V_{2-}) \! \right) \no\\
\end{eqnarray}
where $U_{1-}$ and $U_{2-}$ are both $6D$ spinors with negative chirality and we introduce the short-hand notation:
\begin{displaymath}
\Omega = {[1^\perp 3^\perp]^3 \over [1^\perp 2^\perp][2^\perp 3^\perp] }\ , \qquad
\Omega^\pm =  \Big( 1 \pm {k_1 \cdot q \over k_2 \cdot q} \Big) \Omega \ .
\end{displaymath}
To rewrite the expressions above in a more transparent form, we choose ten-dimensional $\Gamma$ matrices which are listed in Appendix \ref{App-gamma}.
We now need to introduce some additional notation. We label the massive vector fields constructed as the product of two spin-$1/2$ fields as $W_{\mu \xi a}$, where $\mu$ is a spacetime index, $a=1,2,3,4$ is a $SO(5)$ R-symmetry spinor index and $\xi=1,2$ is an additional index labelling the massive gravitino multiplet to which the massive vector field belongs. In addition, $W_{\mu \xi a}$ has a gauge $SU(2)$ fundamental index which is not explicitly displayed. Denoting by $\overline{W}_{\mu}^{\xi a}$ the conjugate field, we have a total of 16 massive vectors belonging to massive gravitini multiplets. They are complemented with two massive vectors $W$ and $\overline{W}$ which belong to two distinct massive vector multiplets and are not charged under the $SU(2)$ part of the gauge group.

Apart from massive vectors, the theory also contains 10 massless vectors.
Vectors constructed combining a scalar field from the explicitly-broken theory with a vector in the spontaneously-broken
theory are  labeled as $A_\mu^4, A_\mu^i$ ($i=7,8,9$). Their indices are inherited from the gauge-theory scalars.  Vectors constructed combining a scalar field from the spontaneously-broken theory with a vector in the explicitly-broken theory are  labeled as $\tilde A_\mu^4, \tilde A_\mu^I$ ($I=5,6,7,8,9$).
Using $\Omega^\pm$ defined above,
the amplitudes involving two massive and one massless vectors with our specific choice of external polarizations are
\begin{eqnarray}
{\cal M}_3\big ( 1 \overline{W}_{+}^{\xi a}, 2  W_{- \zeta b}, 3 A_+^{i} \big ) &= & -i {\kappa \over 2\sqrt{2}} (\lambda/4 )  \delta^\xi_\zeta \delta^a_b  \sigma^{i-6} \  \Omega^- \ , \qquad i=7,8,9  \\
{\cal M}_3\big ( 1 \overline{W}_{+}^{\xi a}, 2  W_{- \zeta b}, 3 A_+^{4} \big ) &= & -i {\kappa \over 2\sqrt{2}} (\lambda/4 )  \delta_\zeta^\xi \delta^a_b \    \Omega^+ \ ,  \\
{\cal M}_3\big ( 1 \overline{W}_{+}^{\xi a}, 2  W_{-\zeta b}, 3 \tilde A_+^4 \big ) &= & -i {\kappa \over 2\sqrt{2}} (\lambda/4 ) \delta_\zeta^\xi \delta^a_b \ \Omega^- \ ,   \\
{\cal M}_3\big (1 \overline{W}_{+}^{\xi a}, 2  W_{-\zeta b}, 3 \tilde A_+^I \big ) &= & -i {\kappa \over 2\sqrt{2}} (\lambda/4 ) \delta_\zeta^\xi  (\tilde \Gamma^{4 I})_b^{a} \ \Omega^+ \ , \qquad I=5,\ldots,9 \ ,
\end{eqnarray}
where the subscripts label the vectors' polarizations. We recall that $\tilde A^{4}_\mu$ is the vector field originating from the scalar acquiring the VEV in the spontaneously-broken gauge theory.
The Pauli matrices $\sigma^{1,2,3}$ act on the gauge $SU(2)$ fundamental indices which are not explicitly displayed.
$\tilde \Gamma^{5I}$ are the R-symmetry generators.
Additional amplitudes involving one or more massless vectors are:
\begin{eqnarray}
{\cal M}_3\big ( 1 A^{i}_+, 2 A^j_-, 3 A^k_+ \big ) &= & \sqrt{2} {\kappa}  (\lambda/4)  \epsilon^{ijk}\Omega \ , \\
{\cal M}_3 \big ( 1 \overline{W}_+, 2 W_-, 3  \tilde A^4_+ \big ) &= & i{\kappa \over 2 \sqrt{2}}  (\lambda/4) (\Omega^- - \Omega^+) \ .
\end{eqnarray}
To further elucidate the meaning of these amplitudes, we recall that the spinor-helicity structures above correspond to the following covariant terms:
\begin{eqnarray}
{\Omega_- \over \sqrt{2} } &=& (p_{\bar W} - p_W) \cdot \varepsilon_A \, (\varepsilon_W \cdot \varepsilon_{\bar W}   ) -  (p_{A} \cdot \varepsilon_{\bar W})  (\varepsilon_A \cdot \varepsilon_{W}   ) + (p_{A} \cdot \varepsilon_W)  (\varepsilon_A \cdot \varepsilon_{\bar W}   ) \no \\
&& \qquad \rightarrow \overline{D_{[\mu} W_{\nu]}}D^{\mu} W^{\nu} \\
{\Omega_+ \over \sqrt{2} } &=& - (p_{A} \cdot \varepsilon_{\bar W})  (\varepsilon_A \cdot \varepsilon_{W}   ) + (p_{A} \cdot \varepsilon_W)  (\varepsilon_A \cdot \varepsilon_{\bar W}   )  \quad \rightarrow \quad   i \overline{W_{\mu}} F^{\mu \nu} W_{ \nu} \ .
\end{eqnarray}
Hence, we see that $A_\mu^{7,8,9}$ become the three $SU(2)$ gluons.
Furthermore, the combination
\begin{equation}
g_{sg} = \kappa \lambda/4
\end{equation}
appears both inside covariant derivatives and field strengths and is identified with the supergravity gauge coupling constant.\footnote{The $SU(2)$ generators in the fundamental representation are normalized as $t^i = {1 \over 2} \sigma^i$.}
The $U(1)$ photon is the linear combination of the vectors $A_\mu^4 = \phi^4 \otimes A_\mu $ and $\tilde A_\mu^4 = A_\mu \otimes \phi^4$ which does not belong to the gravity supermultiplet.

\subsection{Explicitly-broken theories}

\subsubsection{Example with $\cN=4$ unbroken supersymmetry}

In this case, we
start from two copies of the $\cN=2^*$ theory (iii). The gauge group is now taken to be $SU(4N)$.
Denoting with $R(\theta)$ the $U(1)$ transformation acting on the two halves of the hypermultiplet with conjugate phases, we orbifold the first theory as
\begin{equation}
\Phi \rightarrow R\Big({2 \pi \over 3}\Big) g^\dagger \Phi g \ , \qquad
g_{\text{left}} = \text{diag}\big( I_{N}, I_{N},  e^{{4\pi \over 3}i} I_{N}, e^{{2\pi \over 3}i} I_{N} \big) \ . \end{equation}
The second theory is orbifolded as
\begin{equation}
\Phi \rightarrow R\Big({2 \pi \over 3}\Big) g^\dagger \Phi g \ , \qquad
g_{\text{right}} = \text{diag}\big( I_{N}, e^{{2\pi i \over 3}} I_{N}, e^{{4\pi i \over 3}} I_{N} , e^{{4\pi i\over 3}} I_{N} \big) \ .
\end{equation}

For the scalar VEVs for the two theories we take
\begin{eqnarray}
\text{left}: \quad \langle \phi^4 \rangle = {1 \over g} \text{diag}\big( u_1 I_{N} , u_2 I_{N}, u_3 I_{N}, u_4 I_{N} \big) \ , \no \\
\text{right}: \quad \langle \phi^4 \rangle = {1 \over g} \text{diag}\big(\tilde u_1 I_{N_1} , \tilde u_2 I_{N}, \tilde u_3 I_{N}, \tilde u_4 I_{N} \big) \ ,
\end{eqnarray}
with $u_1+u_2+u_3+u_4=0=\tilde u_1+ \tilde u_2+ \tilde u_3+ \tilde u_4$.
Schematically, we get the following organization for the fields of the two theories:
\begin{equation}
\left( \begin{array}{cccc} {\cal V}^{(0)}_{\cN=2} & {\cal V}^{(m)}_{\cN=2} & {\Phi}^+_{\cN=2} & {\Phi}^-_{\cN=2} \\
{\cal V}^{(m)}_{\cN=2}  & {\cal V}^{(0)}_{\cN=2}& {\Phi}^+_{\cN=2} & {\Phi}^-_{\cN=2} \\
{\Phi}^-_{\cN=2} & {\Phi}^-_{\cN=2} &   {\cal V}^{(0)}_{\cN=2}  & {\Phi}^+_{\cN=2} \\
{\Phi}^+_{\cN=2} &  {\Phi}^+_{\cN=2} &  {\Phi}^-_{\cN=2} &   {\cal V}^{(0)}_{\cN=2}     \end{array} \right) \ \ \otimes \ \
\left( \begin{array}{cccc}
{\cal V}^{(0)}_{\cN=2}  & {\Phi}^+_{\cN=2}  & {\Phi}^-_{\cN=2} & {\Phi}^-_{\cN=2} \\
{\Phi}^-_{\cN=2} &   {\cal V}^{(0)}_{\cN=2}  & {\Phi}^+_{\cN=2} &  {\Phi}^+_{\cN=2}  \\
{\Phi}^+_{\cN=2} & {\Phi}^-_{\cN=2} & {\cal V}^{(0)}_{\cN=2} & {\cal V}^{(m)}_{\cN=2} \\
{\Phi}^+_{\cN=2} & {\Phi}^-_{\cN=2} & {\cal V}^{(m)}_{\cN=2}  & {\cal V}^{(0)}_{\cN=2}
\end{array} \right)^t \ .
\end{equation}
As before, some representations need to be combined, which results on constraints on the corresponding VEV parameters. We have the following representation structure for both theories:
\begin{equation}
\left( \begin{array}{cccc}
G & R_1 & R_2 & R_3 \\
\bar R_1 & G & \bar R_3 & \bar R_2 \\
\bar R_2 &  R_3 & G &  R_4 \\
\bar R_3 & R_2 & \bar R_4 & G \\
\end{array} \right) \ . \label{repsN4ex}
\end{equation}
From combining different irreducible representations into the representations $R_2,R_3$ and their conjugates, we get
the following relations between the VEV parameters:
\begin{equation}
u_1 + u_2 = u_3+ u_4 \qquad \rightarrow \qquad  u_4=-u_3 \ , \qquad u_2=-u_1 \ .
\end{equation}
Solving mass-matching conditions leads to\footnote{The choice $\tilde u^2_1 = u^2_1,  \tilde u^2_3 = u^2_3$ also solves the constraints, but does not allow for non-Abelian gauge interactions.}
\begin{equation}
m^2=-M^2 = \tilde m^2 = -\tilde M^2= 4(u_1^2-u_3^2) \ , \qquad \tilde u^2_1 = u^2_3, \qquad \tilde u^2_3 = u^2_1 .
\end{equation}
We list the fields from the double copy with their respective mass spectra in Table \ref{Tab-N4ex}.
The unbroken gauge symmetry is $U(1)^2$, with one $U(1)$ factor from each gauge-theory copy.
The  gauge coupling constant is related to the parameters above as
\begin{equation}
g_s = 2 \kappa \sqrt{u_1^2-u_3^2}  \ .
\end{equation}
In particular, to have a sensible theory we need to take $u_1^2 \geq u_3^2$.
\begin{table}[t]
	\begin{center}
		\begin{tabular}{ccccc}
			Rep. & L & R & sugra fields & mass${}^2$ \\	
			\hline \\
			$G$ & ${\cal V}_{\cN=2}$ & ${\cal V}_{\cN=2}$ & ${\cal H}_{\cN=4} \oplus 2 {\cal V}^{(0)}_{\cN=4} $  & $0$  \\
			$R_1$ & ${\cal V}^{(m)}_{\cN=2}$ & ${\Phi}^+_{\cN=2}$ & ${\Psi}^{(m)}_{\cN=4} $ & $4u_1^2$ \\
			$\bar R_1$ & ${\cal V}^{(m)}_{\cN=2}$ & $ { \Phi}^-_{\cN=2}$ & ${\Psi }^{(m)}_{\cN=4} $ & $4u_1^2$  \\
			$R_4$ &  ${\Phi}^+_{\cN=2}$ & ${\cal V}^{(m)}_{\cN=2}$ & ${\Psi }^{(m)}_{\cN=4} $ & $4u_1^2$  \\
			$\bar R_4$ &  ${\Phi}^-_{\cN=2}$ & ${\cal V}^{(m)}_{\cN=2}$ & ${\Psi }^{(m)}_{\cN=4} $ & $4u_1^2$ \\
			$R_2$ &  ${\Phi}^+_{\cN=2}$ & $ {\Phi}^-_{\cN=2}$ & ${\cal V }^{(m)}_{\cN=4} $  & $(u_1-u_3)^2 + 4u_1^2 - 4 u_3^2 $  \\
			$\bar R_2$ &   ${\Phi}^-_{\cN=2}$ & ${\Phi}^+_{\cN=2}$  & ${\cal V }^{(m)}_{\cN=4} $    & $(u_1-u_3)^2 + 4u_1^2 - 4 u_3^2  $  \\
			$R_3$ &  ${\Phi}^-_{\cN=2}$ & ${\Phi}^-_{\cN=2}$ & ${\cal V }^{(m)}_{\cN=4} $  & $(u_1+u_3)^2 + 4u_1^2 - 4 u_3^2 $ \\
			$\bar R_3$ &   ${\Phi}^+_{\cN=2}$ & ${ \Phi}^+_{\cN=2}$  & ${\cal V }^{(m)}_{\cN=4} $  & $(u_1+u_3)^2 + 4u_1^2 - 4 u_3^2 $
		\end{tabular}
		\caption{Spectrum of the explicitly-broken theory  with $\cN=4$ residual supersymmetry and $U(1)^2$ unbroken gauge group. \label{Tab-N4ex}}	
	\end{center}
\end{table}

\subsubsection{Example with $\cN=0$ unbroken supersymmetry\label{Sec-examples-N0}}

We start from two copies of theory (ii) from Section \ref{Sec-sol}. The gauge group is taken to be $SU(4N)$ as in the previous example. We consider the Coulomb branch of the theory giving a VEV to $\phi^4$ and project away fields non-invariant under the transformation
\begin{equation}
\Phi \rightarrow R_{56}\Big({4\pi \over 3}\Big) g^\dagger \Phi g
 \ , \end{equation}
We take the same  expressions for the $SU(4N)$ elements $g_\text{right}, g_\text{left}$ as in the previous subsection. We further set the scalar mass-matrix to
\begin{displaymath}
m_{55}= m = m_{66} \ , \qquad {m_{IJ}=0} \ \ \text{ otherwise} \ .
\end{displaymath}
The VEVs for the two theories are written as follows,
\begin{eqnarray}
\text{left}: \quad \langle \phi^4 \rangle = \text{diag}\big( u_1 I_N , u_2 I_N, u_3 I_N, u_4 I_N \big) \ , \no \\
\text{right}: \quad \langle \phi^4 \rangle = \text{diag}\big(\tilde u_1 I_N , \tilde u_2 I_N, \tilde u_3 I_N, \tilde u_4 I_N \big) \ ,
\end{eqnarray}
with $u_1+u_2+u_3+u_4=0=\tilde u_1+ \tilde u_2+ \tilde u_3+ \tilde u_4$.
We note that we cannot give VEVs to any other scalar without either breaking the supergravity gauge
group or spoiling the symmetry used for the orbifold projection.
Schematically, we get the following organization for the fields of the two theories:
\begin{equation}
\left( \begin{array}{cccc} A_\mu , \phi^i & W_\mu, \varphi^i & \psi^r , \phi^+ & \tilde \psi^{r'} , \phi^- \\
W_\mu, \varphi^i  & A_\mu , \phi^i& \psi^r , \phi^+ & \tilde \psi^{r'} , \phi^- \\
\tilde \psi^{r'} , \phi^- & \tilde \psi^{r'} , \phi^- &   A_\mu , \phi^i  & \psi^r , \phi^+ \\
\psi^{r} , \phi^+ &  \psi^{r} , \phi^+ &  \tilde \psi^{r'} , \phi^- &   A_\mu , \phi^i     \end{array} \right) \ \ \otimes \ \
\left( \begin{array}{cccc}
A_\mu , \phi^i  & \psi^r , \phi^+  & \tilde \psi^{r'} , \phi^- & \tilde \psi^{r'} , \phi^- \\
\tilde \psi^{r'} , \phi^- &   A_\mu , \phi^i  & \psi^{r} , \phi^+ &  \psi^{r} , \phi^+   \\
\psi^r , \phi^+ & \tilde \psi^{r'} , \phi^- & A_\mu , \phi^i & W_\mu, \varphi^i \\
\psi^r , \phi^+ & \tilde \psi^{r'} , \phi^- & W_\mu, \varphi^i  & A_\mu , \phi^i
\end{array} \right)^t \ .
\end{equation}

As before, some representations need to be combined. The representation structure and mass-matching conditions are the same as in the $\cN=4$ explicitly-broken example.
We list the fields from the double copy with their respective mass spectra in Table \ref{Tab-N0}.\\

\begin{table}[t]
	\begin{center}
		\begin{tabular}{ccccc}
			Rep. & R & L & sugra fields & mass${}^2$  \\	
			\hline \\
			$G$ & $A_\mu \oplus \phi^i$ & $A_\mu \oplus \phi^i$ & $1h_{\mu\nu} \oplus 8 A_\mu \oplus 18 \phi$ & $0$  \\
			$R_1$ & $W_\mu \oplus \phi^i$ & $\psi^r \oplus \phi^+$ & $2 \psi_{\mu} \oplus  W_\mu \oplus 8 \chi \oplus 3 \phi$ & $4u_1^2$  \\
			$\bar R_1 $ & $W_\mu \oplus \phi^i $ & $ \tilde \psi^{r^{\prime}} \oplus \phi^- $ & $2 \psi_{\mu} \oplus  W_\mu \oplus 8 \chi \oplus 3 \phi$ & $4u_1^2 $  \\			
			$R_4$ &  $\psi^r \oplus \phi^+$ & $W_\mu \oplus \phi^i$ &  $2 \psi_{\mu} \oplus  W_\mu \oplus 8 \chi \oplus 3 \phi$ & $4 u_1^2 $  \\		
			$\bar R_4$ &  $\tilde \psi^{r'} \oplus \phi^-$ & $W_\mu \oplus \phi^i$ &  $2 \psi_{\mu} \oplus  W_\mu \oplus 8 \chi \oplus 3 \phi$ & $4 u_1^2 $  \\
						$R_2$ &  $\psi^r \oplus \phi^+$ & $ \tilde \psi^{r'}\oplus \phi^-$ &  $4 W_\mu \oplus 4 \chi \oplus 5 \phi$ & $(u_1-u_3)^2 + 4u_1^2 - 4 u_3^2 $  \\
			$\bar R_2$ &   $ \tilde \psi^{r'}\oplus \phi^-$ & $\psi^r \oplus \phi^+$  &  $4 W_\mu \oplus 4 \chi \oplus 5 \phi$ & $(u_1-u_3)^2 + 4u_1^2 - 4 u_3^2 $  \\
			$R_3$ &  $\tilde \psi^{r'} \oplus \phi^-$ & $ \tilde \psi^{r'}\oplus \phi^-$ &  $4 W_\mu \oplus 4 \chi \oplus 5 \phi$ & $(u_1+u_3)^2 + 4u_1^2 - 4 u_3^2 $   \\
			$\bar R_3$ &   $ \psi^{r}\oplus \phi^+$ & $\psi^r \oplus \phi^+$  &  $4 W_\mu \oplus 4 \chi \oplus 5 \phi$ & $(u_1+u_3)^2 + 4u_1^2 - 4 u_3^2 $
		\end{tabular}
\caption{Spectrum of the explicitly-broken theory with $SO(4)\times U(1)^2$ unbroken gauge symmetry and  no residual supersymmetry.  \label{Tab-N0}}
	\end{center}
\end{table}
Putting all together, we have a two-parameters family of theories with $SO(4)\times U(1)^2$ unbroken gauge symmetry and no residual supersymmetry. The scalar spectrum has three distinct nonzero masses. It should be noted that the mass spectrum does not obey the supertrace conditions identified in \cite{Catino:2013ppa},
\begin{equation}
 \text{Str} {\cal M}^{2k} = 0 \ ,
\end{equation}
where ${\cal M}$ denotes the mass matrix. This fact signals that we are dealing with an explicitly-broken theory.

\section{Conclusion and discussion}
\renewcommand{\theequation}{5.\arabic{equation}}
\setcounter{equation}{0}

We have discussed the construction of Abelian and non-Abelian gaugings of $\cN=8$ supergravity from the double-copy perspective. Our results rely on the existence of several massive deformations of $\cN=4$ SYM theory in higher dimensions which preserve the duality between color and kinematics and possess trilinear scalar couplings. These couplings contain the basic information that specifies the supergravity gauge-group structure constants. In several of the examples we discussed, the  trilinear scalar couplings are accompanied by fermion mass matrices proportional to rank-three elements of the Clifford algebra. In contrast to the perhaps more familiar construction of Higgsed supergravities \cite{Chiodaroli:2015rdg}, these mass terms do not admit a straightforward interpretation as compact higher-dimensional momenta; it remains to be seen whether a different geometrical interpretation can be found.

Our analysis relies on two main tools: spontaneous symmetry breaking and the orbifold projection. Both procedures are known to preserve color/kinematics duality \cite{Chiodaroli:2013upa,Chiodaroli:2015rdg} and yield gauge theories with multiple matter representations which can be used directly in the double-copy construction thus avoiding possibly
cumbersome checks of the duality. At the same time, it is possible that gauge theories  obtained by other means give additional  gaugings of $\cN=8$ supergravity which cannot be constructed with the techniques presented in this paper.

We have also discussed in detail several  examples of gaugings of $\cN=8$ supergravity. They illustrate that our construction can give both supergravities with spontaneously-broken and explicitly-broken supersymmetry. Gaugings with spontaneously-broken supersymmetry can be obtained when one of the gauge theories in the constructions is spontaneously broken. In contrast, the double-copy of two explicitly-broken gauge theories gives a gravitational theory in which supersymmetry is broken by explicit mass terms. It would be interesting to consider double copies of more general spontaneously-broken gauge theories, particularly theories with spontaneously-broken supersymmetry.

The current methods for exploring the non-compact generators of the U-duality group make use of single- and multi-soft scalar
limits \cite{ArkaniHamed:2008gz}.  As we have seen, one of the examples realizes a gauging of $\cN=8$ supergravity
with a noncompact gauge group -- $CSO^*(4,4)\subset E_{7(7)}$. It is possible that the methods developed here may
offer a path towards a double-copy identification of the noncompact U-duality generators as well as a mean for understanding
their implications in the quantum theory.

To understand whether all supergravity theories are double copies, it is important to develop a
direct relation between traditional supergravity methods and the double copy.
While certain properties of the gaugings under consideration (e.g. unbroken gauge group, mass spectrum) can be easily identified
from the double-copy perspective, others remain  elusive. More specifically, a better understanding of the choice of symplectic frame and of the role played by duality transformations would be desirable. It would also be very interesting to see how the embedding tensor can be traced back to the gauge theories entering the construction.

While this paper focuses on gaugings of $\cN=8$ supergravity,
we emphasize that the massive gauge theories introduced in Section \ref{Sec-CK} can also be used as building blocks for extending the double-copy construction to larger families of homogeneous Yang-Mills-Einstein theories, along the lines of ref. \cite{Chiodaroli:2015wal}. More specifically, the solution of the master constraint discussed in Section  \ref{sec-Dsol} can be used to give a construction of Yang-Mills-Einstein theories with arbitrary gauge groups, as long as the number of vector multiplets is large enough. We plan to return to this direction in the future.

A complete classification of double-copy-constructible  gaugings with Minkowski vacua is still missing. Gaugings with large-rank groups appear especially challenging from this perspective. This is because, to this date, trilinear scalar couplings provide the
only known mechanism for introducing non-Abelian gauge interactions in a supergravity through the double copy.
Conversely, vector fields that are realized as bilinears of gauge-theory fermions do not appear to be involved in unbroken gauge interactions. More work in this direction is necessary if a complete classification of gaugings from the double copy is to be obtained.

Our results open the door to a systematic study of the amplitudes of gauged supergravities with $\cN=8$ or less supersymmetry. Given the prominent role played by  ungauged maximal supergravity in recent calculations, it would be natural to investigate the ultraviolet properties of various gaugings with Minkowski vacua. In this respect, the gauging with $\cN=4$ residual supersymmetry appears particularly amenable to loop-level calculations.

\section*{Acknowledgments}

We would like to thank Guillaume Bossard and Henning Samtleben for useful conversations. We also thank Gianguido Dall'Agata for  helpful correspondence. The research of M.C. and H.J. is supported in part by the Knut and Alice Wallenberg Foundation under grant KAW 2013.0235, the Ragnar S\"{o}derberg Foundation under grant S1/16, and the Swedish Research Council under grant 621-2014-5722. The research of R.R. was supported in part by the US Department of Energy under grant DE-SC0013699. M.G. would like to thank the Stanford Institute of Theoretical Physics for its hospitality in the course of this work.

\newpage
\appendix
\section{Spinor-helicity conventions \label{app-conventions}}
\renewcommand{\theequation}{A.\arabic{equation}}
\setcounter{equation}{0}

In this appendix we spell out our conventions. Our notation differs from the one of ref. \cite{ElvangHuang} by the replacement $\eta_{\mu\nu} \rightarrow - \eta_{\mu\nu}$. Our metric has mostly-minus signature both in four and higher dimensions, and the Dirac matrices obey
\begin{equation}
\{ \gamma^\mu , \gamma^\nu \} = 2 \eta^{\mu  \nu} \ , \qquad \{ \Gamma^M , \Gamma^N \} = 2 \eta^{MN} \ .
\end{equation}
In particular, $\gamma^0$ is Hermitian while the other Dirac matrices are anti-Hermitian. The four-dimensional Dirac matrices are
\begin{eqnarray}
\gamma^0 &= &  \sigma^1 \otimes 1 \ , \no \\
\gamma^1 &= & i \sigma^2 \otimes \sigma^1  \ , \no \\
\gamma^2 &= & i  \sigma^2 \otimes \sigma^2  \ , \no \\
\gamma^3 &= & i \sigma^2 \otimes \sigma^3 \ , \no \\
\gamma_5 &= &  \sigma^3 \otimes 1 \ .
\end{eqnarray}
Charge-conjugation and $B$ matrix are
\begin{equation}
C_4 = \sigma^3 \otimes \sigma^2 \ , \qquad
B= \sigma^2 \times \sigma^2 = \left( \begin{array}{cc} 0 & \epsilon_{\dot a \dot b} \\ \epsilon^{ab} & 0 \end{array} \right) \ .
\end{equation}
with $\epsilon^{12}= - \epsilon_{12} = +1$. The charge-conjugation matrix obeys ${ C}_4^t = - {C}_4$ and $(\gamma^\mu)^t = -{C}_4^{-1} \gamma^\mu {C}_4$.
Employing spinor-helicity variables
\begin{equation}
\lambda(p) = \left( \begin{array}{c}
|p ]_a \\ |p\rangle^{\dot a}
\end{array} \right)
\end{equation}
the Majorana condition becomes
\begin{equation}
\lambda^* = B \lambda  \quad \rightarrow \quad \left\{ \begin{array}{c}
\big( |p]_a \big)^* = \epsilon_{\dot a \dot b} |p\rangle^{\dot b} = |p\rangle_{\dot a} \\
\big( |p \rangle^{\dot a} \big)^* = \epsilon^{a b} |p]_{ b} = |p]^{ a} \\
\end{array} \right. \ .
\end{equation}
Note this condition implies that for real momenta we  have
\begin{equation}
([pq])^* = \langle qp \rangle \ .
\end{equation}
Null momenta can be written using spinor-helicity variables in the following way
\begin{equation}
\cancel{p} = - |p\rangle [p| - |p]\langle p| \ .
\end{equation}
We also use repeatedly the identity
\begin{displaymath}
\langle pq \rangle [qp] = 2 p\cdot q \ .
\end{displaymath}
Massless vector polarizations have expressions
\begin{eqnarray}
\cancel{\varepsilon}_+ &=& {\sqrt{2} \over \langle qp \rangle} \big(  |q\rangle [p| + |p]\langle q| \big) \\
\cancel{\varepsilon}_- &=& {\sqrt{2} \over [qp]} \big(  |p\rangle [q| + |q]\langle p| \big)
\end{eqnarray}

For massive momenta, we have
\begin{equation}
k_i = k_i^\perp + {m^2 \over 2 k_i \cdot q  } q =  k_i^\perp - {M^2 \over 2 k_i \cdot q  } q \ .
\end{equation}
The fermionic mass matrix $M$ anticommutes with the four-dimensional momenta.

Using the spinor-helicity formalism, one obtains the identities
\begin{eqnarray}
2 (\epsilon_1^+ \cdot \epsilon_2^-) (\epsilon_3^+ \cdot k_1)  &=& - \sqrt{2}  {k_1 \cdot q \over k_2 \cdot q} \Omega = {1 \over \sqrt{2}} (\Omega_--\Omega_+) \ ,\\
2 (\epsilon_1^+ \cdot \epsilon_2^-) (\epsilon_3^+ \cdot k_1) + \text{cyclic}  &=& \sqrt{2}  \Omega \ ,\\
2 (\epsilon_1^+ \cdot \epsilon_2^+) (\epsilon_3^- \cdot k_1)  &=& 0 \ ,
\end{eqnarray}
where we have used the notation
\begin{displaymath}
\Omega = {[1^\perp 3^\perp]^3 \over [1^\perp 2^\perp][2^\perp 3^\perp] } \ .
\end{displaymath}
Massive spinor polarizations are chosen as:
\begin{eqnarray} &&
u_+ =  \left( \begin{array}{c} U_+ |i^\perp ] \\ - M  U_+ {|q\rangle \over \langle i^\perp q \rangle} \end{array} \right),\quad
u_- =  \left( \begin{array}{c} - M U_- {|q ]\over [ i^\perp q ]} \\ U_- |i^\perp \rangle \end{array} \right) , \\
&& \bar v_+ =  \left( \begin{array}{cc} U_-^\dagger [i^\perp | & - U_-^\dagger  {\langle q | \over \langle i^\perp q \rangle} M \end{array} \right)  \ , \\
&& \bar v_- =  \left( \begin{array}{cc}   -U_+^\dagger {[q | \over [i^\perp q]} M & U_+^\dagger \langle i^\perp | \end{array} \right)  \ ,
\end{eqnarray}
here $U_\pm$ are four component Weyl spinors with positive/negative $6D$ chirality. Imposing the Majorana condition in $10D$ yields
\begin{equation}
U_\pm = C_6^{*} U_\mp^* \ ,
\end{equation}
where $C_6$ is the charge-conjugation matrix in the compact directions. Using this relation, we see that
our choice of polarization satisfies
\begin{equation}
u_\pm = B_{10}^{-1} (u_\mp)^*\ .
\end{equation}

\section{Gamma matrices\label{App-gamma}}
\renewcommand{\theequation}{B.\arabic{equation}}
\setcounter{equation}{0}

To study the gauging with $\cN=4$ residual supersymmetry encountered in Section \ref{ex-N4},  we choose ten-dimensional Dirac matrices in a chiral basis. The expressions for the spacetime matrices are:
\begin{eqnarray}
\gamma^0 &= &  \sigma^1 \otimes 1 \ , \no \\
\gamma^1 &= & i \sigma^2 \otimes \sigma^1  \ , \no \\
\gamma^2 &= & i  \sigma^2 \otimes \sigma^2  \ , \no \\
\gamma^3 &= & i \sigma^2 \otimes \sigma^3 \ , \no \\
\gamma_5 &= &  \sigma^3 \otimes 1 \ .
\end{eqnarray}
While the Dirac matrices in the internal directions are:
\begin{eqnarray}
\Gamma^4 &= & i \gamma_5 \otimes \sigma^2 \otimes \sigma^3 \otimes \sigma^1 \ , \no \\
\Gamma^5 &= & i \gamma_5 \otimes \sigma^2 \otimes \sigma^3 \otimes \sigma^2\ ,  \no \\
\Gamma^6 &= & i \gamma_5 \otimes \sigma^2 \otimes \sigma^3 \otimes \sigma^3 \ , \no \\
\Gamma^7 &= & i \gamma_5 \otimes \sigma^1 \otimes 1 \otimes 1 \ , \no \\
\Gamma^8 &= & i \gamma_5 \otimes \sigma^2 \otimes \sigma^1 \otimes 1 \ , \no \\
\Gamma^9 &= & i  \gamma_5 \otimes \sigma^2 \otimes \sigma^2 \otimes 1 \ .
\end{eqnarray}
The corresponding expressions for charge-conjugation and chirality matrix are
\begin{equation}
C_4 = \sigma^3 \otimes \sigma^2  \ , \qquad  C_6 =  \sigma^2 \otimes \sigma^1 \otimes \sigma^2 \ , \qquad \Gamma^*_6 = \sigma^3 \otimes 1 \otimes 1 \ .
\end{equation}
Note that the four-dimensional basis of gamma matrices is the same as in Elvang and Huang \cite{ElvangHuang}. We also have
\begin{equation}
C_{10}= C_4 C_6 \ ,\qquad B_{10}= \sigma^2 \otimes \sigma^2 \otimes \sigma^2 \otimes \sigma^1 \otimes \sigma^2 \ .
\end{equation}
This choice satisfies the relations
\begin{equation}
C_{10} \Gamma^A C_{10}^{-1} = -  (\Gamma^A)^t \ , \qquad B_{10} \Gamma_{*} B^{-1}_{10} =  (\Gamma_{*})^* \ ,
\end{equation}
where $\Gamma_*$ is the ten-dimensional chirality matrix (the last condition is necessary for the existence of Majorana-Weyl spinors).
We note that the $M$ matrix can be written as
\begin{equation}
M = i {\lambda \over 4} \Gamma^{456} = i {\lambda \over 4} \gamma_5 (\sigma^2 \otimes \sigma^3 \otimes 1) \ .
\end{equation}
Similarly, the generator for rotations in the $5$-$6$ plane (used in the orbifold projection) has a diagonal form:
\begin{equation}
\Gamma^{89} = i 1 \otimes \sigma^3 \otimes 1 \ .
\end{equation}

Using
\begin{equation}
R_i = e^{\phi_i \Gamma^{89}}
\end{equation}
The requirement (\ref{CPT-cond}) becomes
\begin{equation}
(C_6 e^{\phi_i \Gamma^{89}} C_6^{-1} ) (\Lambda_G G \Lambda_G) = 1 \ .
\end{equation}
Remembering that $G$ acts as an overall phase on the various unbroken group representations and that $(\Lambda_G G \Lambda_G) = G^{-1}$, we can finally write
\begin{equation}
(C_6 e^{\phi_i \Gamma^{89}} C_6^{-1} ) = e^{-\phi_i \Gamma^{89}}
\end{equation}
which is indeed satisfied for our choice of Dirac matrices.

\section{Feynman rules}
\renewcommand{\theequation}{C.\arabic{equation}}
\setcounter{equation}{0}

We conclude by listing some of the Feynman rules to be employed in the calculations. The fermionic propagator is written as
\begin{equation}
\begin{array}{c} \includegraphics[scale=0.6]{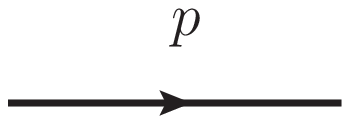} \\ \ \end{array} = -i {\cancel p + M \over p^2 + M^2} .
\end{equation}
Here we take
	$M$ to be a matrix which anticommutes with $\cancel{p}$. Note that this expression is different from the one of standard Quantum Field Theory books,  $-i {\cancel p - M \over p^2 - M^2}$, which is derived under the assumption that $M$ is proportional to the identity.
Note that $M$ is a matrix in the flavor indices, which squares to a  diagonal matrix but is otherwise generic.
Scalar and gluon propagators are
\begin{equation}
 \begin{array}{c} \includegraphics[scale=0.6]{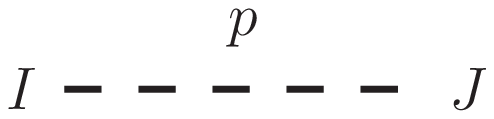}  \end{array} \hskip -0.5cm = \ \ i {\delta^{IJ} \over p^2 } , \qquad \qquad
\begin{array}{c} \includegraphics[scale=0.6]{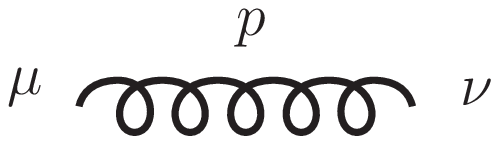} \end{array} \hskip -0.5cm = \ \  -i {\eta^{\mu \nu} \over p^2 } .
\end{equation}
We list only the vertices relevant to the computations in this paper:
\bea
 \begin{array}{c} \\ \includegraphics[scale=0.6]{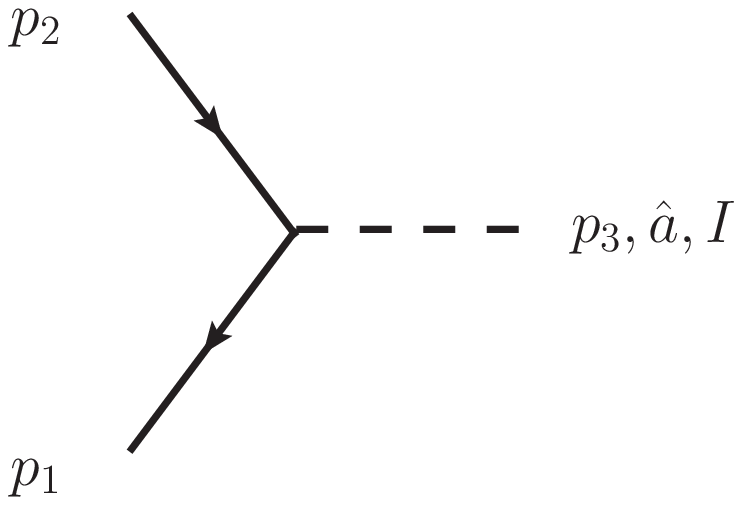} \end{array} \hskip -0.2cm &=&  i g \Gamma^I   t^{\hat a} \ ,
 \eea
 \bea
 \begin{array}{c} \\ \includegraphics[scale=0.6]{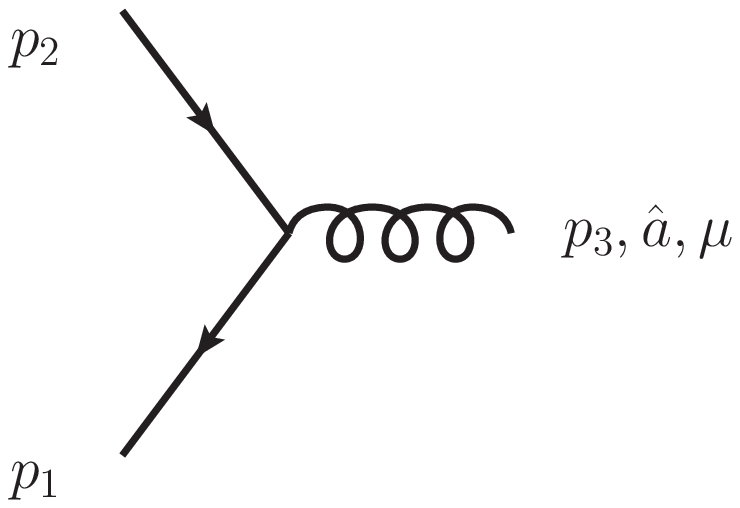} \end{array} \hskip -0.2cm  &=& i g \gamma^{\mu} t^{\hat a} \ , \\
 \begin{array}{c} \\ \includegraphics[scale=0.6]{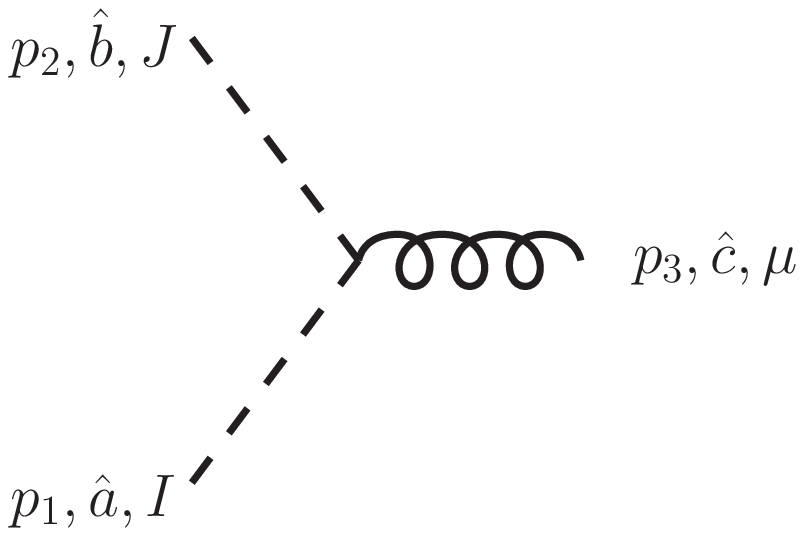} \end{array} \hskip -0.2cm  &=& g  f^{\hat a \hat b \hat c} \delta^{IJ} (p_1-p_2)^\mu \ , \\
\begin{array}{c} \\ \includegraphics[scale=0.6]{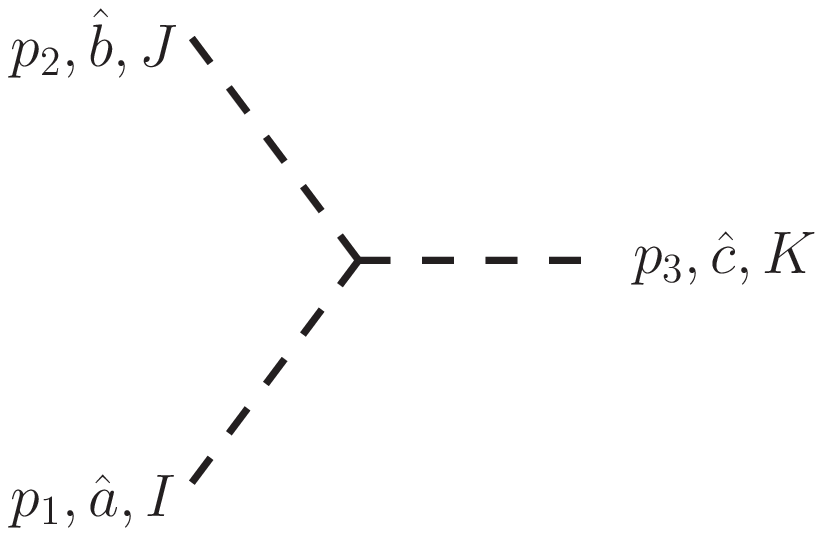} \end{array} \hskip -0.2cm &=& - i g \lambda  f^{\hat a \hat b \hat c} F^{IJK} \ .
\eea
These rules are employed for studying the gauge-theory amplitudes in Section \ref{Sec-CK}.

\end{document}